\documentclass[12pt]{article}
\usepackage{amsmath} \usepackage{amssymb}
\usepackage{bm}
\usepackage{bbm}
\usepackage{cite}
\usepackage[dvips]{color}
\usepackage{graphicx}
\usepackage{slashed} \usepackage{tabularx}

\newcommand{\note}[1]{#1} 
\newcommand{\GeV}{\mbox{ GeV}}

\newcommand{\cm}{\mbox{ cm}}

\begin{document}
\title{\Large\bf
  Implications of the latest XENON100 and  cosmic ray  antiproton data for
  isospin violating dark matter
}

\vspace{0.3truecm}
\author{
  Hong-Bo Jin\footnote{Email: hbjin@itp.ac.cn},
  Sen Miao\footnote{Email: miaosen@itp.ac.cn}
  and  Yu-Feng Zhou\footnote{Email: yfzhou@itp.ac.cn}
  \\ \\
 \textit{State Key Laboratory of Theoretical Physics},\\
  \textit{Kavli Institute for Theoretical Physics China,}\\
  \textit{Institute of  Theoretical Physics, Chinese Academy of Sciences}\\
  \textit{Beijing, 100190, P.R. China}
}
\date{}
\maketitle

\begin{abstract}
  In the scenario of isospin violating dark matter (IVDM), the dark matter
  (DM) spin-independent couplings to protons and to neutrons are allowed to be
  different, which has been considered to relax the tensions between the
  results of DAMA, CoGeNT and XENON experiments. We explore the allowed values
  of DM-nucleon couplings favored and excluded by the current experiments
  under the assumption of IVDM. We find that the recently updated XENON100
  result excludes the main part of the  overlapping signal region between
  DAMA and CoGeNT.     We also show that the possible tensions between some experiments such
  as that between DAMA and SIMPLE are unlikely to be affected by isospin
  violating interactions.  In an effective operator approach, we investigate
  conservative upper bounds on the DM-quark couplings required by the IVDM
  scenario from the cosmic ray antiproton fluxes measured recently by
  BESS-Polar II and PAMELA, and that from the relic density.  The results show
  that the relatively large couplings favored by DAMA and CoGeNT are tightly
  constrained, if the operators contribute to velocity-independent
  annihilation cross sections.  For thermal relic DM, the upper bounds from
  the relic density can also be stringent.     \end{abstract}

\newpage

\section{Introduction}

It has been well established from astrophysical and cosmological observations
that nearly 85$\%$ of the matter in the Universe consists of invisible dark
matter (DM). So far the evidence of DM arise solely from its gravitational
interactions. Popular DM candidates, such as the weakly interacting massive
particles (WIMPs) can naturally reproduce the observed relic density through weak
interactions with the standard model (SM) particles. This possibility has
motivated numerous experiments to probe the direct or indirect signals of DM
interacting with ordinary matter.

Some of the recent DM direct detection experiments such as DAMA
\cite{0804.2741, 0808.3607,1002.1028}, CoGeNT \cite{1002.4703,1106.0650} and
CRESST-II \cite{1109.0702} have reported events in excess of known
backgrounds. The excess events, if interpreted in terms of DM particle elastic
scattering off target nuclei, may imply light DM particles with mass around
8-10 GeV and scattering cross section around $10^{-40}\cm^{2}$. Other
experiments such as CDMS-II \cite{1010.4290, 1011.2482}, XENON10/100
\cite{1104.3088,1104.2549}, and SIMPLE \cite{1106.3014} etc., have reported
null results in the same DM mass range. The understanding of  the backgrounds and
systematic uncertainties in the current experiments  still needs to be improved.
At the moment, It is premature to draw a conclusion based on a single
 experimental result.

A commonly adopted assumption on interpreting the DM direct detection data is
that in spin-independent scatterings the DM particle couplings to proton
$(f_{p})$ and to neutron $(f_{n})$ are nearly the same, i.e. $f_{n}\approx
f_{p}$, which makes it straight forward to extract the DM-nucleon scattering
cross sections.  However, in generic
cases, the interactions may be isospin-violating
~\cite{hep-ph/0307185,hep-ph/0504157,1003.0014,1004.0697,1102.4331,1105.3734,1112.6364}.
In the scenario of isospin violating DM (IVDM), the DM particle couples to
proton and neutron with different strengths, possible destructive interference
between the two couplings can weaken the bounds from XENON10/100 and move the
signal regions of DAMA and CoGeNT to be closer to each other
\cite{1102.4331,1105.3734}. In order to reconcile the data of DAMA, CoGeNT and
XENON10/100, a large destructive interference corresponding to $f_{n}/f_{p}\approx -0.7$ is
favored \cite{1102.4331}.

Recently, the XENON100 collaboration has reported updated results based on the
exposure of 225 days with 34 kg fiducial mass \cite{1207.5988}. Two events are
observed with background expectation of $1.0\pm0.2$ events ($0.79\pm0.16 $
from gammas and $0.17^{+0.12}_{-0.07}$ from neutrons).  The updated upper
bound on the spin-independent scattering cross section is $2\times 10^{-45}
\text{ cm}^{2}$ at 90$\%$ confidence level (CL) for a 55 GeV DM particle,
which is improved by a factor of 3.5 in comparison with the previous result
reported in 2011 \cite{1104.2549}.  For an 8 GeV DM particle, the limit is
improved by a factor of five. Such a significant improvement may allow the
XENON100 result to challenge the DAMA and GoGeNT results even in the scenario
of IVDM.

As a consequence of  $f_{n}/f_{p}\approx -0.7$,  the absolute values $|f_{p,n}|$ have to be around two order
of magnitudes larger than that obtained under the assumption of IC
interaction. At quark level, this means that the absolute DM couplings to the
first generation $u$- and $d$-quarks must be enhanced by the same order of
magnitudes.
The scenario of IVDM with such light DM masses and relatively large couplings
to light quarks can be tested in other processes in which there is no
destructive interference.  For instance, the light DM particles can be pair
produced at the Large Hadron Collider (LHC) and the Tevatron with signals of a
single jet/photon plus a large missing transverse energy, i.e. a
mono-jet/mono-photon.  The current null search results from LHC
\cite{1109.4725,CMS-monojet} and Tevatron \cite{CDF-monojet} can be used to
impose constraints on the DM-quark couplings
\cite{1005.3797,1005.1286,1008.1783,1204.3839} and the IVDM \cite{1108.1196}.

As the inverse processes of DM production, light DM particles can annihilate
into light-quark pairs in the galactic halo, which provides exotic sources of
cosmic gamma rays, neutrinos and antiprotons. The annihilation processes occur
at low velocities $v/c \approx \mathcal{O}(10^{-3})$ where $c$ is the speed of
light, which can be used to test the IVDM complementary to that at the LHC and
Tevatron.  For such low energy processes, the effective operator approach
can be valid as long as the intermediate mediator particles are
significantly heavier than the DM particle, but could be much lighter than the
TeV scale (for recent discussions, see e.g. Refs. \cite{1005.1286,1008.1783,1009.0008,1010.1774,1011.2310,1107.3529}).
Possible constraints from the cosmic neutrinos and gamma ray on IVDM have been
discussed in Refs. \cite{ 1103.3270,1106.4044,1112.4849}. Recently the
BESS-Polar II experiment has measured the antiproton flux in the energy range
from 0.2 GeV to 3.5 GeV \cite{1107.6000} which have  higher precision compared
with that from PAMELA  \cite{1103.2880} at low energies.  The antiproton
flux in this energy range can receive significant contributions from the annihilation
of 10 GeV scale DM particles, provided that the
annihilation cross section is not velocity suppressed.
The constraints from BESS-Polar II data have been investigated recently for
typical thermal WIMP using semi-analytical approach \cite{1110.4376}.  The
constraints on the IVDM was briefly discussed in
Ref. \cite{1112.4849} which however did not provide  a detailed analysis on the model parameters and uncertainties.
In this work, we shall perform a more systematic  analysis on  
implications of the recent cosmic ray antiproton flux data for IVDM,
which is based on the fully numerical GALPROP approach. We shall compare
several propagation models and aim at obtaining  conservative upper bounds on the isospin-violating
couplings.

For thermal relic DM, the same annihilation process also determines the DM
relic density. By requiring that the calculated relic density from a single
annihilation channel should not be smaller than the observed value $\Omega
h^{2}=0.113\pm 0.004$ \cite{1001.4538}, upper bounds on the DM couplings can
be obtained.  The bounds could be stringent if the annihilation is dominated
by $s$-wave processes.  For $p$-wave annihilation, although the cross section
is velocity-suppressed, useful upper bounds can still be obtained as the
typical relative velocity of DM particles is finite $v/c \approx 0.3$ at
freeze out,

In this work, We first explore the values of DM-nucleon couplings favored by
the current experiment under the assumption of IVDM for various target
material and values of $f_{n}/f_{p}$.  We find that the recently updated
XENON100 result is able to rule out  the main part of the  marginally  overlapping signal region
between DAMA and CoGeNT \note{unmodulated data} for both the cases with and without
considering the surface event rejection factors.
 We also show that the tensions between some group of experiments
are unlikely to be affected by isospin violation, especially that between DAMA
and SIMPLE.  We adopt the effective operator approach to investigate the
constraints from the antiproton flux data and the thermal relic density on the
couplings between the IVDM and the SM light quarks. We calculate the
antiproton flux using the
numerical method implemented in the GALPROP code
\cite{astro-ph/9807150,astro-ph/0106567,astro-ph/0101068,astro-ph/0510335},
and consider a number of propagation parameter configurations in order to obtain
conservative upper bounds.
The results  show that the large and
negative value $f_{n}/f_{p}=-0.7$ is severely constrained for the operators
with velocity-independent annihilation cross sections, such as the fermionic
DM with vector  couplings and complex scalar DM with scalar
couplings.  For these operators we also find that the constraints from the
cosmic antiproton flux  are stronger than that from the thermal relic
density.  The relic density can provide useful constraint in the case where
the operators contribute only to velocity suppressed annihilation cross
sections. For the complex scalar DM with derivative couplings, we find that
the IVDM is in some tension with the relic density.

This paper is organized as follows. In Section \ref{sec:IVDM}, we discuss the
effect of IVDM on various target nuclei in different  experiments and then explore the allowed DM couplings
to nucleons and to light quarks. In Section \ref{sec:effective-interactions},
we list the effective operators relevant to the IVDM and the related interaction cross sections.  In Section
\ref{sec:pbar}, we derive  constraints on IVDM from the current cosmic ray
antiproton data. In  section \ref{sec:relic-density} the constraints from the DM relic
density is discussed.  The numerical results are presented in Section \ref{sec:results} and
the conclusions are given in Section \ref{sec:conclusions}.

\section{Isospin violating dark matter}\label{sec:IVDM}
For a DM particle $\chi$ with mass $m_{\chi}$ elastically scattering off a
target nucleus with atomic number $Z$ and atomic mass number $A$, the recoil
event rate $R=dN/dt$ is given by
\begin{align}\label{eq:7}
R=N_{T}\left( \frac{\rho_{0}}{m_{\chi}}\right)\int dE_{R} \int_{v_{min}}^{v_{esc}}d^{3}v f(v) v \frac{d\sigma}{dE_{R}} \ ,
\end{align}
where $N_{T}$ is the total number of the target nuclei and $\rho_{0}$ is the local DM energy density \note{which takes
typical values in the range $0.2-0.56\text{ GeV}\cdot\text{cm}^{-3}$ \cite{1107.5810}}.  For a fixed recoil energy
$E_{R}$, the required minimal velocity of the DM particle is
$v_{min}=[m_{A}E_{R}/(2\mu_{A}^{2})]^{1/2}$, where $m_{A}$ is the mass of the target nucleus and
$\mu_{A}=(m_{\chi}m_{A})/(m_{\chi}+m_{A})$ is the DM-nucleus reduced
mass.  The maximal velocity is the
escape velocity from our Galaxy at the position of the Solar system
$v_{esc}\approx 600 \mbox{ km}\cdot\mbox{s}^{-1}$.  The differential
scattering cross section can be written as
\begin{align}\label{eq:9}
  \frac{d\sigma}{dE_{R}}=
  \frac{m_{A}F^{2}(E_{R})}{2\mu_{A}^{2} v^{2}}\sigma_{0}  \ ,
\end{align}
where $F(E_{R})$ is the form factor of the nucleon. Here we have assumed that
the form factors for proton and neutron are nearly identical, i.e.,
$F_p(E_R)\approx F_n(E_R) \equiv F(E_{R})$.  The quantity $\sigma_{0}$ can be understood as
the total scattering cross section at the limit of zero-momentum transfer which
is related to $f_{p(n)}$  through
\begin{align}\label{eq:21}
\sigma_{0}=\frac{\mu_{A}^{2}}{\pi}\left[Zf_{p} +(A-Z)f_{n}\right]^{2} \ .
\end{align}
\note{Here it is assumed that the cross section is independent of the velocity of the DM particle.} 
The  cross section for the DM particle scattering off a free nucleon
in term of $f_{p(n)}$ is
\begin{align}
\label{eq:47}
\sigma_{p(n)}=\frac{\mu_{p(n)}^{2}}{\pi} f_{p(n)}^{2},
\end{align}
where $\mu_{p(n)}$ is the DM-proton (neutron) reduced mass.
Under the assumption that the scattering is isospin conserving (IC),
i.e., $f_{n}\approx f_{p}$, the total cross section $\sigma_{0}$ is independent of
$Z$ and only  proportional to $A^{2}$.  One can define a
cross section $\sigma_{p}^{IC}$ which is the value of $\sigma_{p}$ extracted  from
$\sigma_{0}$ under the assumption of IC interaction as
\begin{align}\label{eq:19}
\sigma_p^{IC} \equiv \frac{\mu_{p}^{2}}{\mu_{A}^{2}A^{2}} \sigma_{0} \ ,
\end{align}
which is the quantity  commonly reported  by the experiments.
In the generic case where $f_{n} \neq f_{p}$, the true value of $\sigma_{p}$ will
differ from $\sigma^{IC}_{p}$ by a factor $F(f_n/f_p)$ which depends on the
ratio $f_n/f_p$ and the target material
\begin{align}\label{eq:19}
\sigma_p = F(f_n/f_p) \sigma_p^{IC} .
\end{align}
Depending on the mass of the DM particle, for a given target material,
for instance  $\mbox{CaWO}_{4}$ used by CRESST-II experiment
there could be multiple-target  nuclei
relevant to the nuclear recoil. If  the target material consists of  $N$ kind of relevant
nuclei with atomic numbers $Z_{\alpha} \ (\alpha=1,\dots, N)$ and
fractional number abundances  $\kappa_{a}$,  and for each nucleus $Z_{\alpha}$
there exists  $M$ type of isotopes found in nature with atomic mass number
$A_{\alpha i}$ and fractional number abundance $\eta_{\alpha i}
\ (i=1,\dots,M)$, the expression of $F(f_{n}/f_{p})$ can be explicitly written as
\begin{align}
\label{eq:34}
F(f_{n}/f_{p})=
\frac{\sum_{\alpha,i}\kappa_{\alpha}\eta_{\alpha i} \mu_{A_{\alpha i}}^{2}A_{\alpha i}^{2}}{\sum_{\alpha, i}\kappa_{\alpha}\eta_{\alpha i}\mu_{A_{\alpha i}}^{2}[Z_{\alpha}+(A_{\alpha i}-Z_{\alpha})f_{n}/f_{p}]^{2}} \ ,
\end{align}
where $\mu_{A_{\alpha i}}$ is the reduced mass for the DM and the nucleus  with
atomic mass number $A_{\alpha i}$. In the simplest case where the target consists of one kind of
nucleus $(Z, A)$,  $F(f_n/f_p)$ takes the  simple form
\begin{align}\label{eq:19}
F(f_{n}/f_{p})= \left[ \frac{Z}{A}+\left(1-\frac{Z}{A} \right) \frac{f_{n}}{f_{p}}\right]^{-2} \ .
\end{align}
It is evident that $F(f_{n}/f_{p})$ approaches unity in the case of IC scattering.  If
$f_{n}/f_{p}<0$ the interference between the contributions from proton and
neutron scattering to the value  of $F(f_n/f_p)$ will be  destructive, which can  lead  to
$F(f_{n}/f_{p}) \gg 1$. Thus  it is possible that the value of $\sigma_{p}$
can be a few order of magnitudes larger than $\sigma^{IC}_{p}$, provided a
nearly complete cancellation between the two contributions.

For a given target material $T$, there is a particular value of $f_{n}/f_{p}$ which corresponds
to the maximal possible value of $F(f_n/f_p)$
\begin{align}
\label{eq:44}
\xi_{T}\equiv
=-\frac{\sum_{\alpha,i} \kappa_{\alpha}\eta_{\alpha i} \mu_{A_{\alpha i}}^{2}(A_{\alpha i}-Z_{\alpha})Z_{\alpha}}{\sum_{\alpha, i} \kappa_{\alpha}\eta_{\alpha i} \mu_{A_{\alpha i}}^{2}(A_{\alpha i}-Z_{\alpha})^{2}} \ .
\end{align}
For a  single nucleus target with atomic (mass) number $Z(A)$, it
is simply given by $\xi_{Z}=-Z/(A-Z)$.  The value of $\xi_{T}$ varies  with
target material. In Tab. \ref{tab:xi}, we list the values of
$\xi_{T}$ for some typical material utilized by the current or future experiments.
\begin{table}[htb]
  \begin{center}
    \begin{tabular}{ccccccccc}\hline
      & Xe     &Ge     & Si  &Na(I)     &Ca(W)$\mbox{O}_{4}$  & $\mbox{C}_{2}\mbox{ClF}_{5}$ & CsI  & Ar\\
      \hline
      $\xi_{T}$& -0.70  &-0.79 & -1.0 &-0.92(-0.73) &-1.0(-0.69)                                     &-0.92                                                &-0.71 &-0.82\\
      \hline
    \end{tabular}
    \caption{Values of $\xi_{T}$ for different target material. For NaI, the two values -0.92 and -0.73 correspond to the scattering off
      Na and NaI respectively. Similarly, for $\text{CaWO}_{4}$, the two values  -1.0 and  -0.69 corresponds to the scattering without and
    with tungsten nuclei respectively.}  \label{tab:xi}
  \end{center}
\end{table}

\begin{figure}[htb]
\begin{center}
  \includegraphics[width=0.45\textwidth]{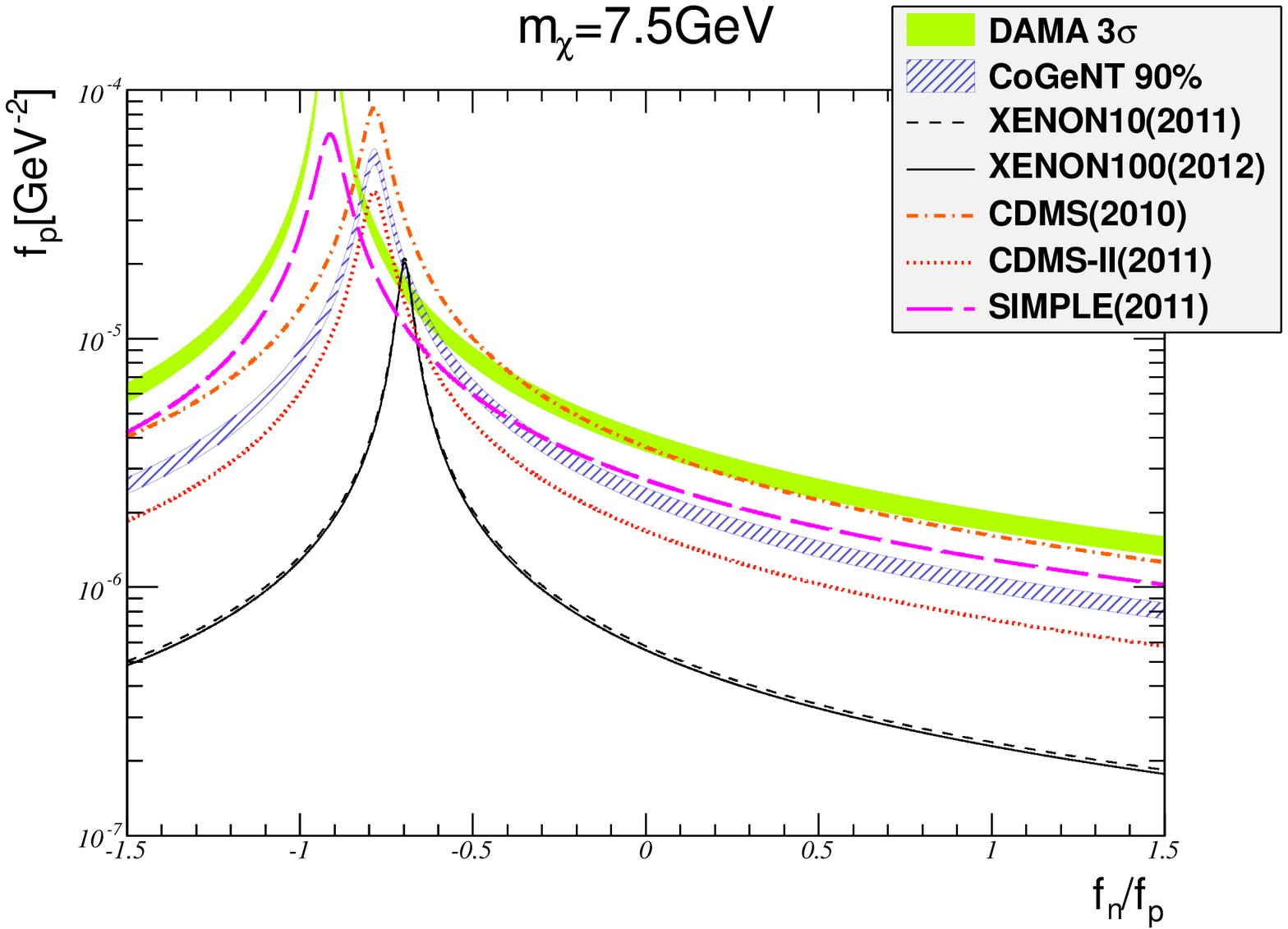}
  \includegraphics[width=0.45\textwidth]{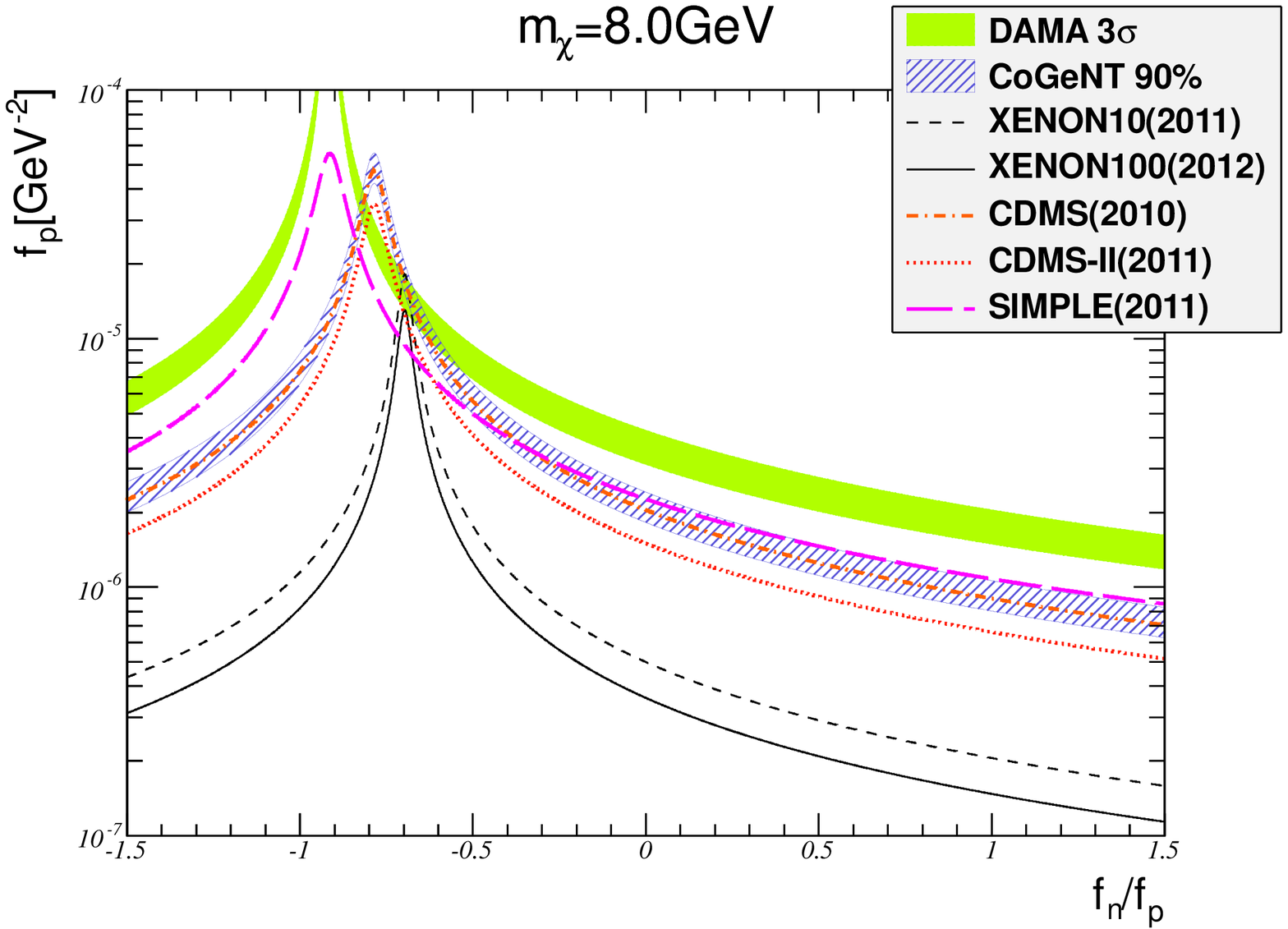}\\
  \includegraphics[width=0.45\textwidth]{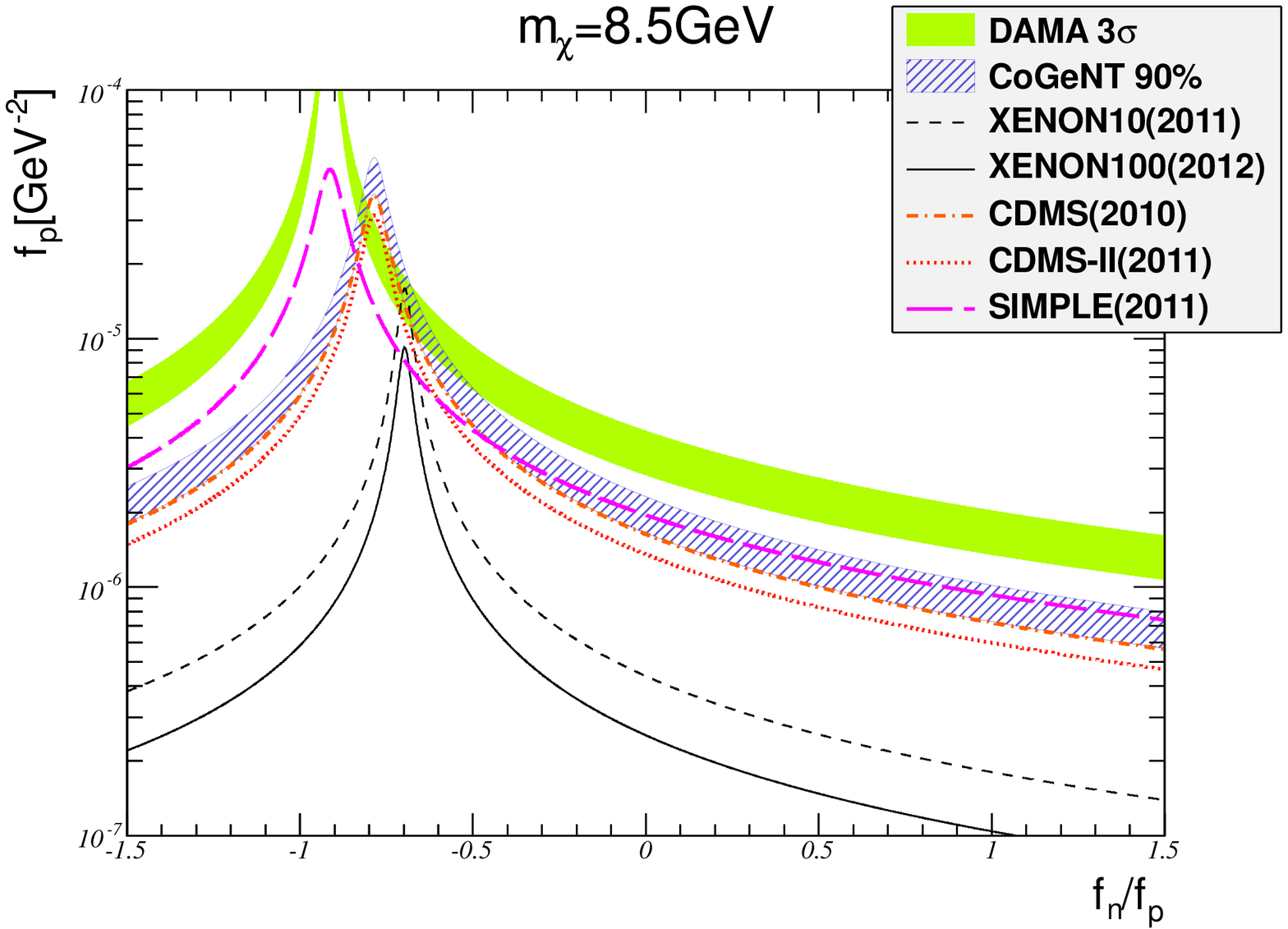}
  \includegraphics[width=0.45\textwidth]{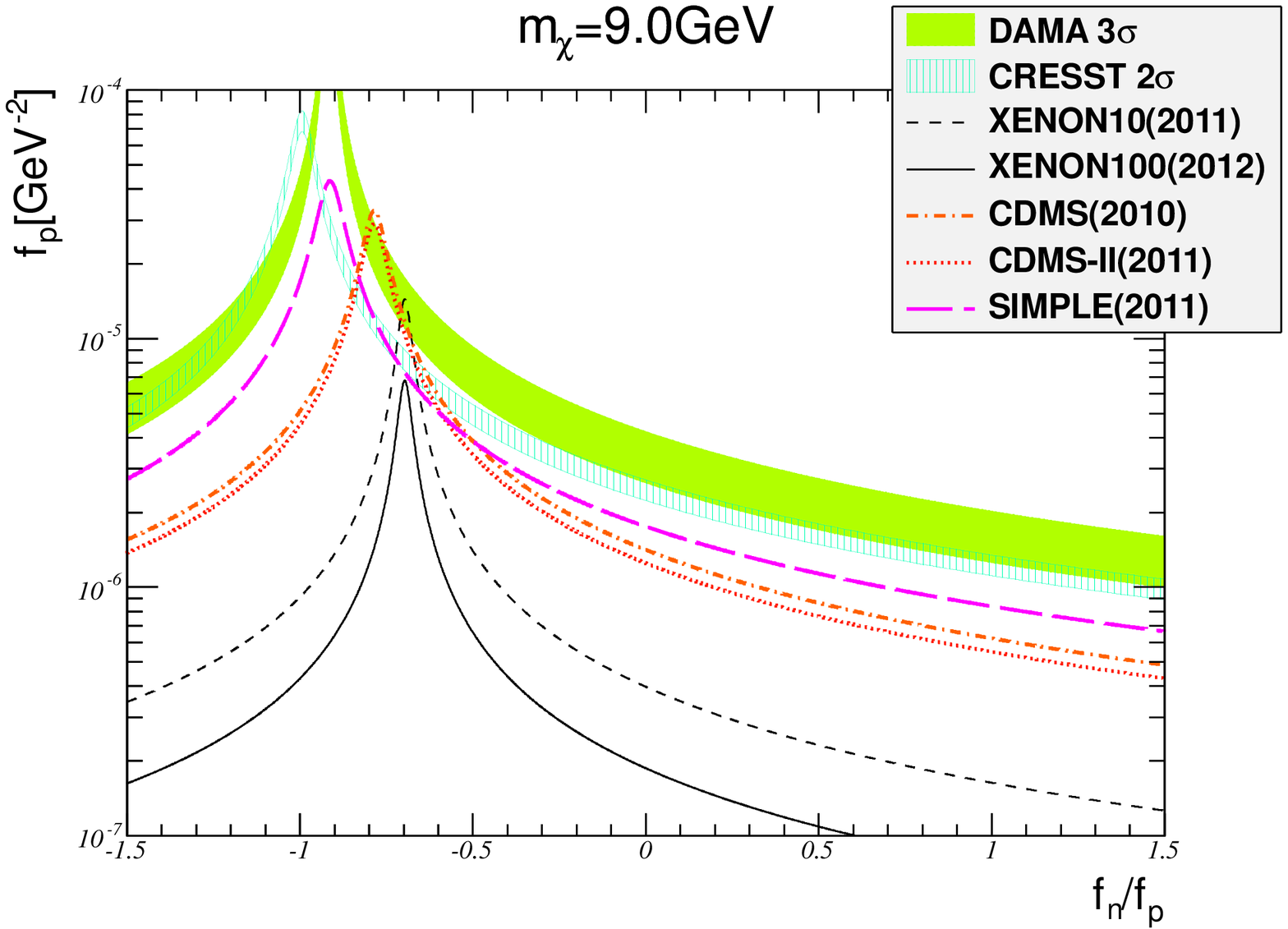}\\
\end{center}
\caption{Favored regions and  upper bounds  in ($f_{p},f_{n}/f_{p}$) plane  from current experiments
 DAMA \cite{0808.3607},
  GoGeNT \cite{1106.0650}, XENON \cite{1104.3088,1207.5988}, CDMS \cite{1010.4290,1011.2482} and SIMPLE \cite{1106.3014} are also shown. Four panels corresponds to the four different mass of dark matter
  particle fixed at 7.5 GeV, 8.0 GeV, 8.5 GeV and 9.0 GeV respectively.
}
\label{fig:fnfp}
\end{figure}

In Fig. \ref{fig:fnfp}, we plot the regions favored and excluded by the
current experiments in the plane of $(f_p, f_n/f_p)$ for four different DM masses from 7.5 GeV to 9.0 GeV.
The experimental results
include:
90$\%$ CL  upper bounds from CDMS-II Si and Ge\cite{1010.4290,1011.2482},
CRESST-II $2\sigma$ favored region for scattering without involving tungsten nucleus \cite{1109.0702},
DAMA $3\sigma$ favored region assuming Na scattering without considering  the channeling effects \cite{1002.1028},
\note{ GoGeNT $90\%$ favored region without considering the surface rejection correction factor \cite{1106.0650}},
90 $\%$ CL upper bound from SIMPLE \cite{1106.3014},
the 90$\%$ CL upper bound from XENON10  which is insensitive to scintillation efficiency \cite{1104.3088} and
the 90$\%$ CL upper bound from the XENON100 \cite{1207.5988}.
As it has been noticed previously at $f_n/f_p=-0.70=\xi_{\text{Xe}}$, the DAMA- and CoGeNT-favored regions can overlap,
the sensitivities of XENON10/100 are maximally reduced by two order of magnitudes. However, the DAMA- and
CoGeNT favored regions  are only marginally consistent with the new XENON limit  for $m_{\chi}=7.5$ and 8.0 GeV.
For $m_{\chi}=8.5$ and 9.0 GeV, the overlapping regions start to disappear. \note{The CRESST-II result favors a heavier DM particle. The
  allowed region   by CRESST-II at $2\sigma$ can only be seen in the lower-right frame in Fig. \ref{fig:fnfp} for $m_{\chi}=9.0$ GeV, and is
below the DAMA-favored region.}

\begin{figure}[htb]
\begin{center}
  \includegraphics[width=0.60\textwidth]{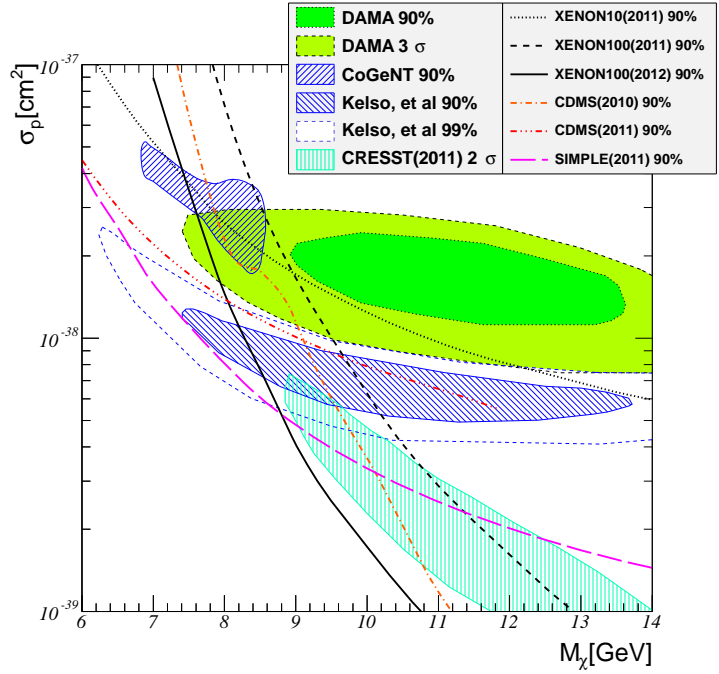}
\end{center}
\caption{The favored regions and constraints in the $(\sigma_p,
  m_{\chi})$ plane for various experiments for $f_{n}/f_{p}=-0.70$ such as
 DAMA \cite{0808.3607},
  GoGeNT unmodulated data \cite{1106.0650},  CoGeNT unmodulated data with surface event rejection factors taken from Kelso, etal \cite{1110.5338}, CRESST-II \cite{1109.0702}, XENON10/100 \cite{1104.3088,1207.5988}, CDMS \cite{1010.4290,1011.2482} and SIMPLE \cite{1106.3014}. See text for explanation.
}
\label{fig:direct_IV}
\end{figure}

\note{  In Fig. \ref{fig:direct_IV}, the allowed regions by the current experiments are shown in the
  $(\sigma_{p},m_{\chi})$ plane  for $f_{n}/f_{p}=-0.70$ . In the
  plot, we include the CoGeNT allowed regions at $90\%$ and $99\%$ CL with the
  surface event rejection correction factor taken from Ref. \cite{1110.5338}.
  The DAMA-favored  region at $90\%$ CL is also shown \cite{1002.1028}.  When
  the corrections from the surface event rejection are taken into account, the
  GoGeNT-favored region moves towards  larger DM mass and  lower cross
  section. As it can be seen, for $f_{n}/f_{p}=-0.70$, the corrected
  GoGeNT-favored region corresponds to $m_{\chi}\approx 10$ GeV and
  $\sigma_{p}\approx 10^{-38}\cm^{2}$ which has marginal overlap with both
  DAMA- and CRESST-favored regions. Very recently, the CoGeNT has reported 
  updated correction factors \cite{1208.5737} which are a little bit higher
  than that used in Ref. \cite{1110.5338}, thus the GoGeNT-favored region may
  move closer to the signal region of DAMA.}

\note{
  As shown in  Fig. \ref{fig:direct_IV}, at $f_{n}/f_{p}=-0.70$, the overlapping
  region between GoGeNT and DAMA may still be consistent with the exclusion curve
  from the XENON100 2011 data \cite{1104.2549}. However,  If  one considers  the recently  updated upper bounds
  from  XENON100 \cite{1207.5988}, the main bulk of the overlapping region is excluded
  for both the GoGeNT results with and without surface event rejection corrections.
  Thus the recent XENON100 2012 result has a significant  impact on the understanding of  the nature of DM.
  It challenges the IVDM as a scenario to reconcile the results of DAMA, CoGeNT and XENON.
}

\note{
  From Fig. \ref{fig:direct_IV}, at $f_{n}/f_{p}=-0.70$, the overlapping region
  between DAMA and CoGeNT seems also to be excluded by the results of SIMPLE
  \cite{1106.3014} and CDMS-II independently \cite{1010.4290,1011.2482}.  Note that there exists 
  controversies regarding the detector stability of SIMPLE experiments
  \cite{1106.3559,1107.1515}, the recoil energy calibration of CDMS experiment
  \cite{1103.3481} and the extrapolation of the measured scintillation
  efficiency to lower recoil energy in the previous XENON100 data analysis
  \cite{1005.0838,1005.2615}. The recently updated XENON100 result not only
  has considerably larger exposure and substantial reduction of background
  from $^{85}\text{Kr}$, but also adopts a hard cut in S1 acceptance, which  removes
  nearly all the event below the measured scintillation efficiency and 
  makes the extrapolation to low recoil energy irrelevant.   }

If the $\xi_{T}$ values of the target material used by two experiments are
very close to each other, the tension between the two experimental results, if
exists, is less affected by the effect of isospin violation. From
Tab. \ref{tab:xi} one finds that $\xi_{\text{Na}}\approx
\xi_{\text{C}_2\text{ClF}_5}=-0.92$, $\xi_{\text{Xe}}\approx
\xi_{\text{CsI}}\approx -0.7$ and $\xi_{\text{Si}}\approx
\xi_{\text{Ca(W)O}_{4}}=-1.0$.
Thus the tension between DAMA
signal from Na recoil and the upper bound from SIMPLE is unlikely to be alleviated by
isospin violation, which can be clearly seen in Fig.
\ref{fig:fnfp}.  Similarly, if there exists
contradictions  between XENON and KIMS, CoGeNT and the Ar based experiments
such as  DarkSide, it  can hardly be  explained by isospin violating scattering.
 The SIMPLE result is also useful in
comparing with the CRESST-II which utilizes Ca(W)$\mbox{O}_{4}$ which has
$\xi_{\text{Ca(W)O}_{4}}=-1.0$.
Obviously, for the experiments use the same target material, the possible
tension between them cannot be relaxed by isospin violation, such as the
tension between CoGeNT and CDMS-II, as both use germanium  as target nucleus.

\section{Effective interactions}\label{sec:effective-interactions}
We assume that the DM particles interact with the SM light quarks through some
heavy mediator particles much heavier than the DM particle
such that both the scattering and the annihilation processes can be  effectively
described by a set of high dimensional contact operators
\begin{align}
\label{eq:36}
\mathcal{L}=\sum_{i,q} a_{iq}\mathcal{O}_{iq} \ .
\end{align}
If the DM particles are Dirac fermions, the  relevant operators arising  from  scalar  or pseudoscalar  interactions are given by
\begin{align}\label{eq:1}
\mathcal{O}_{1q}= \bar{\chi}\chi\bar{q}q,  \quad
\mathcal{O}_{2q}=\bar{\chi}\gamma^{5}\chi\bar{q}q, \quad
\mathcal{O}_{3q}=\bar{\chi}\chi\bar{q}\gamma^{5}q, \quad
\mathcal{O}_{4q}=\bar{\chi}\gamma^{5}\chi\bar{q}\gamma^{5}q  .
\end{align}
The operators from vector or axial-vector  type interactions are
\begin{align}\label{eq:2}
\mathcal{O}_{5q}= \bar{\chi}\gamma^{\mu}\chi\bar{q}\gamma_{\mu}q, \
\mathcal{O}_{6q}=\bar{\chi}\gamma^{\mu}\gamma^{5}\chi\bar{q}\gamma_{\mu}q, \
\mathcal{O}_{7q}=\bar{\chi}\gamma^{\mu}\chi\bar{q}\gamma_{\mu}\gamma^{5}q, \
\mathcal{O}_{8q}=\bar{\chi}\gamma^{\mu}\gamma^{5}\chi\bar{q}\gamma_{\mu}\gamma^{5}q ,
\end{align}
and the ones from the tensor  interactions are
\begin{align}\label{eq:3}
\mathcal{O}_{9q}=\bar{\chi}\sigma^{\mu\nu}\chi\bar{q}\sigma_{\mu\nu}q,  \quad
\mathcal{O}_{10q}=\bar{\chi}\sigma^{\mu\nu}\gamma^{5}\chi\bar{q}\sigma_{\mu\nu}q .
\end{align}
If the DM particles are Majorana particles the vector and tensor operators are
vanishing. Among these operators only $\mathcal{O}_{1q}$ and
$\mathcal{O}_{5q}$ contribute to spin-independent scattering cross sections at
low velocities.  The scattering cross sections induced by the operators
$\mathcal{O}_{2q}$ and $\mathcal{O}_{6q}$ are velocity suppressed. The
operators $\mathcal{O}_{7q}$ and $\mathcal{O}_{8q}$ contribute only to
spin-dependent scattering cross section, and the nucleus matrix elements for
the operators $\mathcal{O}_{3q}$, $\mathcal{O}_{4q}$, $\mathcal{O}_{9q}$ and
$\mathcal{O}_{10q}$ are either vanishing or negligible.
The DM annihilation into quarks through $\mathcal{O}_{1q}$ is a $p$-wave
process, which is velocity suppressed. It does not contribute to the cosmic
antiproton flux, but  still contributes to  the DM relic density as $p$-wave processes
is non-negligible at freeze out.

Similarly, for DM being a complex scalar $\phi$,  possible operators are
\begin{align}\label{eq:5}
\mathcal{O}_{11}=2m_{\phi}(\phi^{*}\phi) \bar{q}q , \
 \mathcal{O}_{12}=2m_{\phi}(\phi^{*}\phi) \bar{q} \gamma^{5}q , \
\mathcal{O}_{13}=(\phi^{*}\overleftrightarrow{\partial_{\mu}}\phi )\bar{q} \gamma^{\mu} q , \
\mathcal{O}_{14}=(\phi^{*}\overleftrightarrow{\partial_{\mu}}\phi )\bar{q} \gamma^{\mu} \gamma^{5}q .
\end{align}
Among those only $\mathcal{O}_{11}$ and $\mathcal{O}_{13}$ contribute to the
spin-independent scatterings. The DM annihilations through operator
$\mathcal{O}_{13}$ are $p$-wave processes.

The fermionic or scalar DM may couple to gluons through the operators
$\bar{\chi}\chi G_{\mu\nu}G^{\mu\nu}$, $\bar{\chi}\gamma^{5}\chi
G_{\mu\nu}G^{\mu\nu}$, $\bar{\chi}\chi G_{\mu\nu}\tilde{G}^{\mu\nu}$ and
$\bar{\chi}\gamma^{5}\chi G_{\mu\nu}\tilde{G}^{\mu\nu}$. We do not consider
these operators as they do not contribute to isospin violating  scatterings. In summary,
we only consider the four operators
$$
\mathcal{O}_{1q}, \  \mathcal{O}_{5q}, \  \mathcal{O}_{11q}, \mbox{ and } \mathcal{O}_{13q} ,
$$
which are relevant to IVDM.

The DM couplings to nucleons $f_{p,n}$ can be expressed in terms of  the DM couplings to quarks $a_{iq}$ as follows
\begin{align}
\label{eq:37}
f_{p(n)}=\sum_{q} B^{p(n)}_{iq} a_{iq} .
\end{align}
For the Dirac DM with scalar interaction $a_{1q} \bar{\chi}\chi \bar{q}q$, one has  $B^{p(n)}_{1q}=f^{p(n)}_{Tq}\ m_{p(n)}/m_{q} $ for $q=u,d,s$
and $B^{p(n)}_{1q}=(2/27)f^{p(n)}_{TG} m_{p(n)}/m_{q}$ for $q=c,b,t$,
where $f^{p(n)}_{Tq}$ is the DM coupling to light quarks obtained from the  $\sigma$-term $\left\langle N|m_{q}\bar{q}q|N\right\rangle =f_{Tq}^{N}M_{N}$,
and $
f^{p(n)}_{TG}=1-\sum_{q=u,d,s}f^{p(n)}_{Tq}$.  In
numerical calculations we take $f^p_{Tu}=0.020\pm0.004$,
$f^p_{Td}=0.026\pm0.005$, $f^p_{Ts}=0.118\pm0.062$, $f^n_{Tu}=0.014\pm0.003$,
$f^n_{Td}=0.036\pm0.008$ and $f^n_{Ts}=0.118\pm0.062$~\cite{hep-ph/0001005}.
 Using the following quark masses:
$m_{d}=0.005$ GeV, $m_{u}/m_{d}=0.55$, $m_{s}=0.095$ GeV,$m_{c}=1.25$ GeV, $m_{b}=4.2$ GeV and $m_{t}=172.3\mbox{ GeV}$,  we obtain the values of $B_{iq}$:
\begin{align}
\label{eq:39}
& B_{1u}^{p} \simeq B_{1d}^{n}\simeq 6.8, \ B_{1d}^{p}\simeq B_{1u}^{n}\simeq 4.8,
\nonumber\\
& B_{1s}^{p,n} \simeq1.2, \ B_{1c}^{p,n}\simeq 0.05,
\ B_{1b}^{p,n}\simeq 1.5\times 10^{-2}, \mbox{ and }
B_{1t}^{p,n}\simeq 3.5\times 10^{-4} .
\end{align}
In order to maximize the isospin violating effect, the coefficients
$B_{1s,1c,1b,1t}^{p,n}$ must be strongly suppressed.  Assuming that the
DM-nucleon couplings are dominated by the DM couplings to the first generation
quarks, the ratio $f_{n}/f_{p}$ is given by
\begin{align}
\label{eq:40}
\frac{f_{n}}{f_{p}} \approx \frac{B^{n}_{1u}a_{1u}+B^{n}_{1d} a_{1d}}{B^{p}_{1u}a_{1u}+B^{p}_{1d} a_{1d}} .
\end{align}
The value of  $f_{n}/f_{p}=-0.7$ can be translated into  $a_{1d}/a_{1u}= -0.93$ at quark level. This  value is the same for complex scalar DM.
For operator  $\mathcal{O}_{5q}$ one simply has
$B^{p(n)}_{5u}=2(1)$  and $B^{p(n)}_{5d}=1(2)$,  and  $B^{p(n)}_{q}=0$  for $q=c,s,t,b$.
\begin{align}
\label{eq:41}
\frac{f_{n}}{f_{p}}=\frac{a_{5u}+2  a_{5d}}{2 a_{5u}+ a_{5d}} .
\end{align}
Thus for $f_{n}/f_{p}=-0.7$, one finds $a_{5d}/a_{5u}=-0.89$.
 The cross section for DM annihilating into quarks induced by the relevant operators are given by
\begin{align}
\label{eq:48}
\sigma_{1q} v_{\text{rel}}(\bar{\chi}\chi\rightarrow\bar{q}q) &=\frac{N_{C}m_{\chi}^{2} }{2\pi}a_{1q}^{2} v_{\text{rel}}^{2} ,
\nonumber\\
\sigma_{5q}v_{\text{rel}}(\bar{\chi}\chi\rightarrow\bar{q}q)&=\frac{N_{C}m_{\chi}^{2}  }{\pi} a_{5 q}^{2} ,
\nonumber\\
\sigma_{11q}v_{\text{rel}}(\bar{\phi}\phi\rightarrow\bar{q}q)& =\frac{N_{C}m_{\phi}^{2}}{\pi} a_{11q}^{2} ,
\nonumber\\
\sigma_{13q}v_{\text{rel}}(\bar{\phi}\phi\rightarrow\bar{q}q)&=\frac{2N_{C} m_{\phi}^{2}}{3\pi}a_{13q}^{2} v_{\text{rel}}^{2},
\end{align}
where $N_{C}=3$ is the number of color and $v_{\text{rel}}$ is the relative velocity of annihilating DM particles.
For $s$-wave annihilation, the thermally averaged value of the cross section multiplied by velocity $\langle \sigma v_{\text{rel}}\rangle$ is
the same as $\sigma v_{\text{rel}}$.

\section{Cosmic antiproton flux}\label{sec:pbar}

Annihilation or decay of light DM particles in the galactic halo can
contribute to exotic primary sources of the low energy cosmic ray antiprotons,
which can be probed or constrained by the current satellite- and balloon-borne
experiments such as PAMELA and BESS-polar II, etc..  The predicted antiproton
flux from DM annihilation depends on models of the cosmic-ray transportation,
the distribution of Galactic gas, radiation field and magnetic field, etc.. It also
depends on the particle and nuclear interaction cross sections.

In this work, we use the numerical code GALPROP
\cite{astro-ph/9807150,astro-ph/0106567,astro-ph/0101068,astro-ph/0210480,astro-ph/0510335} which utilizes realistic astronomical
information on the distribution of interstellar gas and other data as input
and consider various kinds of data including primary and secondary nuclei,
electrons and positrons, $\gamma$-rays, synchrotron and radiation etc. in a
self-consistent way. Other approaches based on simplified assumptions on the
Galactic gas distribution which  allows  for fast  analytic solutions can be found in
Refs.~\cite{astro-ph/0103150,astro-ph/0212111,astro-ph/0306207,1001.0551}.
In the GALPROP approach, we consider several diffusion models (parameter
configurations). The different results between the models can be regarded as an
estimate of theoretical uncertainties.

In the diffusion models of cosmic ray propagation, the Galactic halo where
diffusion occurs is parameterized by a cylinder with half height $Z_{h}$ and
radius $R=20-30$ kpc.  The densities of cosmic ray particles are vanishing at the
boundary of the halo. The processes of energy losses, reacceleration and
annihilation take place in the Galactic disc. The source terms for the
secondary cosmic rays are also confined within the  disc.
The diffusion  equation for the cosmic ray particle is given by
\begin{align}\label{eq:28}
  \frac{\partial \psi}{\partial t} =&
  \nabla (D_{xx}\nabla \psi -\mathbf{V}_{c} \psi)
  +\frac{\partial}{\partial p}p^{2} D_{pp}\frac{\partial}{\partial p} \frac{1}{p^{2}}\psi
  -\frac{\partial}{\partial p} \left[ \dot{p} \psi -\frac{p}{3}(\nabla\cdot \mathbf{V}_{c})\psi \right]
  \nonumber \\
  & -\frac{1}{\tau_{f}}\psi
  -\frac{1}{\tau_{r}}\psi
  +q(\mathbf{r},p)  ,
\end{align}
where $\psi(\mathbf{r},p,t)$ is the number density per unit of total particle
momentum which is related to the phase space density $f(\mathbf{r},p, t)$ as
$\psi(\mathbf{r},p,t)=4\pi p^{2}f(\mathbf{r},p,t) $. For steady-state diffusion, it is assumed that  $\partial  \psi/\partial t=0$.
The spatial diffusion coefficient $D_{xx}$ is parameterized as
\begin{align}
\label{eq:29}
D_{xx}=\beta D_{0} \left( \frac{\rho}{\rho_{0}} \right)^{\delta}  ,
\end{align}
where $\rho=p/(Ze)$ is the rigidity of cosmic ray particle and $\delta$ is the power spectral index which may  take
different values $\delta_{1}$ or $ \delta_{2}$ when $\rho$ is below or above the reference rigidity
$\rho_{0}$.  $D_{0}$ is a normalization constant, and $\beta=v/c$ is the velocity of the cosmic ray particle.
The values of $D_{0}$ and $\delta$ are determined from the
measurements of the fluxes of other cosmic ray species such as the ratio of Boron to Carbon (B/C) and of isotopes
$^{10}\text{Be}/^{9}\text{Be}$.
The convection term is related to the drift of antiproton from the Galactic
disc due to the Galactic wind.  The direction of the wind is usually assumed to be along the $z$-direction which is
perpendicular to the galactic disc and is a constant
$\mathbf{V}_{C}=[2\theta(z)-1] V_{c}$.
The  diffusion in momentum space is described by the reacceleration parameter $D_{pp}$ which is
related to the   Alfv$\grave{\mbox{e}}$n speed $V_{a}$ of the disturbances in the hydrodynamical plasma as
\begin{align}
\label{eq:30}
D_{pp}=\frac{4V_{a}^{2} p^{2}}{3D_{xx}\delta(4-\delta^{2})(4-\delta)w} ,
\end{align}
where $w$ stands for the level of turbulence which can be taken as $w=1$ as $D_{pp}$ depends only on the
 combination $V_{a}^{2}/w$.
 $\dot{p}$ is related to  the momentum loss rate,
$\tau_{f}$ and $\tau_{r}$ are the time scale for fragmentation and radioactive
decay respectively.

The sources of the primary particles are chosen to reproduce the cosmic-ray
distribution determined by the EGRET $\gamma$-ray data.  The injection spectrum of
nuclei  is assumed to have a broken power low behavior $dq(p)/dp\approx
\rho^{-\gamma}$, with $\gamma=\gamma_{1}(\gamma_{2})$ for the rigidity $\rho$ below (above)
a reference rigidity $\rho_{s}$.
The secondary source is given in terms of the distributions of the primary
particles $\psi(\mathbf{r},p)$ and the distribution of the interstellar medium (ISM)
\begin{align}
\label{eq:33}
q(\mathbf{r},p)=
\beta c \psi_{prim}(\mathbf{r},p) [\sigma_{\text{H}}(p) n_{\text{H}}(\mathbf{r})+\sigma_{\text{He}}n_{\text{He}}(\mathbf{r})] ,
\end{align}
where $n_{\text{H}}(\mathbf{r})$ and $n_{\text{He}}(\mathbf{r})$ are the
number densities of interstellar hydrogen and helium
respectively. $\sigma_{\text{H}}$ and $\sigma_{\text{He}}$ are the cross
sections for the generation of the secondary particles from the interactions
with H and He.  The detailed calculations of the cross sections and the
distribution of interstellar gas can be found in Ref. \cite{astro-ph/0106567}
and references therein.

Thus the whole diffusion process depends on a number of parameters: $R$,
$Z_{h}$, $\rho_{0}$, $D_{0}$, $\delta_{1}/\delta_{2}$, $V_{c}$ , $V_{a}$,
$\rho_{s}$ and $\gamma_{1}/\gamma_{2}$. There exists degeneracies in the
determination of the diffusion parameters from B/C.  In general, increasing
the diffusion half-height $Z_{h}$ can be compensated by the increase of the
diffusion constant $D_{0}$, which leads to significant uncertainties in the
predictions of antiproton flux from DM annihilation as the flux depends
strongly on $Z_{h}$. 
The recently updated analysis based on
Markov Chain Monte Carlo fits to the astrophysical  data show that $Z_{h}$ should be around
$4-7\text{ kpc}$ \cite{1001.0551,1011.0037}.  In order to obtain a
conservative upper bounds we choose $Z_{h}\approx 4 \text{ kpc}$ in our
analysis.

We consider several typical propagation models in GALPROP, and focus on the
models with the secondary antiproton background below than the current data,
which leaves room for DM contribution and results in conservative upper
bounds. The first one is the plain diffusion model (referred to as ``Plain'')
in which there is no reacceleration term \cite{astro-ph/0510335}. The second
one is the conventional model (referred to as ``Conventional'') with
reacceleration included \cite{astro-ph/0101068,astro-ph/0510335}. The last one
(referred to as ``Global-Fit'') is the model from a global fit to the relevant
astrophysical observables using Markov-Chain Monte Carlo method
\cite{1011.0037}. The main parameters of the three models are listed in
Tab. \ref{tab:prop-models}.
Note that in the three models the convection is not considered, as the inclusion of conversion term leads to problems in reproducing
the spectrum of B/C and discontinuity  in the propagation across the Galactic disc in GALPROP\cite{astro-ph/0510335}.
\begin{table}[htb]
  \begin{center}
  \begin{tabular}{lllllllll}\hline
    model &$R (\mbox{kpc})$& $Z_{h}(\mbox{kpc})$ & $D_{0}$ & $\rho_{0}$&$\delta_{1}/\delta_{2}$   & $V_{a}(\mbox{km}/\mbox{s})$ &$\rho_{s}$&$\gamma_{1}/\gamma_{2}$ \\
 \hline
 Plain                &30   &  4.0  & 2.2   &3  &    0/0.60       & 0    &40 &2.30/2.15\\
Conventional               &20   & 4.0   & 5.75 &4  & 0.34/0.34     & 36  & 9  & 1.82/2.36 \\
Global-Fit        &20   & 3.9   & 6.59 &4  &  0.3/0.3        &39.2 &10 & 1.91/2.40\\
  \hline
  \end{tabular}\end{center}
\caption{Propagation parameters in the ``Plain'' \cite{astro-ph/0510335}, ``Conventional'' \cite{astro-ph/0101068,astro-ph/0510335} and
  ``Global-Fit'' \cite{1011.0037} models used in the GALPROP code. $D_{0}$ is in units of  $10^{28}\mbox{cm}^{2}\cdot\mbox{s}^{-1}$, the break rigidities $\rho_{0}$ and $\rho_{s}$  are in units of GV.}
 \label{tab:prop-models}
\end{table}

The primary source term from  the DM annihilation has the form
\begin{align}\label{eq:ann-source}
q(\mathbf{r})=\eta n(\mathbf{r})^{2}\langle \sigma v_{\text{rel}}\rangle \frac{dN}{dp} ,
\end{align}
where $n(\mathbf{r})=\rho(\mathbf{r})/m_{\chi}$, $\eta=1/2(1/4)$ if the DM
particle is (not) its own antiparticle and $dN/dp$ is the injection spectrum
per DM annihilation.
For the DM profile we took the isothermal profile
\cite{astro-ph/9712318}
\begin{align}
\label{eq:32}
\rho(\mathbf{r})=\rho_{0}\left( \frac{r_{\odot}^{2}+R_{s}^{2}}{r^{2}+R_{s}^{2}} \right) ,
\end{align}
where $r_{\odot}=8.5\text{ kpc}$ is the distance of Solar system from the
galactic center, $R_{s}=2.8\text{ kpc}$ and the local density is taken to be
$\rho_{0}=0.3\GeV\cm^{-3}$.  The choice of isothermal profile and the local
density is again to achieve a conservative estimate of the antiproton flux, if
the NFW profile is used, the predicted flux will be enhanced roughly by at most
$70\%$, thus more severe constraints are expected.

The total antiproton  flux is related to the density function as
\begin{align}\label{eq:38}
\Phi= \frac{v}{4\pi} \psi(p) .
\end{align}
In the force-field approximation \cite{Gleeson:1968zza},
the antiproton flux at the top of the atmosphere of the Earth $\Phi^{TOA}_{\bar p}$ which is measured by the
experiments is related to the interstellar antiproton flux as
\begin{align}
\label{eq:45}
\Phi^{TOA}_{\bar p}(T_{TOA})=\left(\frac{2m_{p} T_{TOA}+T_{TOA}^{2}}{2m_{p} T+T^{2}}\right)\Phi_{\bar p}(T) ,
\end{align}
where $T_{TOA}=T-\phi_{F}$ is the antiproton kinetic energy at the top of the atmosphere of the Earth. The BESS-Polar II data were
taken in a period of lowest solar activity. We take $\phi_{F}=0.5$ GV in numerical analysis.

\section{DM thermal relic density}\label{sec:relic-density}

In the case where the DM particles are  thermal relics, possible large couplings to light quarks may
under predict the  DM relic abundance in comparison with the observed  value $\Omega
h^{2}=0.113\pm0.004$ \cite{1001.4538}. Thus the relic density can also impose upper
bounds on the relevant couplings.  Whereas the $p$-wave annihilation processes
give no contribution to the antiproton flux, in the calculation of DM relic
density, the $p$-wave annihilation is nonnegligible  as  the
typical relative velocity  of DM particles is $v_{\text{rel}}/c \approx
\mathcal{O}(0.3)$  at freeze out. The annihilation cross section times the relative velocity
can be expressed  as $\sigma v_{\text{rel}}=a +b v_{\text{rel}}^{2}$, where $a$ and $b$ are coefficients
corresponding to the $s$-wave and $p$-wave contributions. The thermally
averaged value at the time of freeze out has the form $\left\langle \sigma
  v_{\text{rel}}\right\rangle _{f} \simeq a+6b/x_{f}$, where $x_{f}\equiv m_{DM}/T_{f}$
with $T_{f}$ the decoupling temperature. The value of $x_{f}$ can be estimated
through the iterative solution of the equation \cite{hep-ph/0404175}
\begin{equation}
  x_{f}=\ln
  \left[
    C(C+2)\sqrt{\frac{45}{8}}
    \frac{g}{2\pi^{3}}
    \frac{M_{pl}m_{DM}(a+6b/x_{f})}{g_{*}^{1/2}x_{f}^{1/2}}
    \right]  ,
\end{equation}
where $g$ is the number of degrees of freedom of the DM particle, $g_{*}$ is
the number of effective relativistic degrees of freedom of the thermal
plasma and  $M_{pl}=1.22\times 10^{19}\GeV$  is the Planck energy scale.  The
constant $C$ is determined by matching the semi-analytical solutions to the
fully numerical solutions to the Boltzmann equation for the thermal evolution
of particle number density.  We take $C(C+2)=1(2)$ for $s(p)$-wave annihilation
in numerical calculations.  For light DM around 10 GeV, $g_{*}=61.75$, we find $x_{f}\approx 22$.  The relic
abundance is approximately  given by
\begin{align}
\label{eq:46}
\Omega h^{2} \simeq \frac{1.07\times 10^{9} \mbox{ GeV}^{-1}}{M_{pl}}\frac{x_{f}}{\sqrt{g_{*}}}
\frac{1}{a+3b x_{f}} .
\end{align}

\section{Results}\label{sec:results}
We perform $\chi^{2}$ analysis to obtain upper bounds on the DM isospin
violating couplings to light quarks. Since within uncertainties the current
BESS-Polar II and PAMELA data can be consistent with the secondary antiproton
background, we consider one-side exclusion limit in which only the data points
below the theoretical prediction are taken into account, namely we calculate
the quantity $\chi^{2}=\sum_{i}
(\Phi^{\text{th}}_{i}-\Phi^{\text{exp}}_{i})^{2}/\sigma_{i}^{2}$, ( if
$\Phi^{\text{th}}_{i}>\Phi^{\text{exp}}_{i}$), where $\Phi^{\text{exp}}_{i}$
are the measured fluxes with uncertainty $\sigma_{i}$ and
$\Phi^{\text{th}}_{i}$ are the theoretical predictions at kinetic energy
$T_{i}$, and set upper bounds at $95\%$ CL. The required
$\chi^{2}$ values for different degrees of freedom (d.o.f) are calculated from
the standard $\chi^{2}$-distribution. For instance, the required $\chi^{2}$
values are $64$, $49.8$, and $31.4$ for d.o.f=47, 35, and 20 respectively. The
upper bounds obtained in this manner is more conservative than that using the whole
data set. The constraints from the relic density are also set at $95\%$ CL.
For the sake of simplicity, we consider one operator at a time and ignore
interference between the operators.

\begin{figure}[htb]
\begin{center}
 \includegraphics[width=0.60\textwidth]{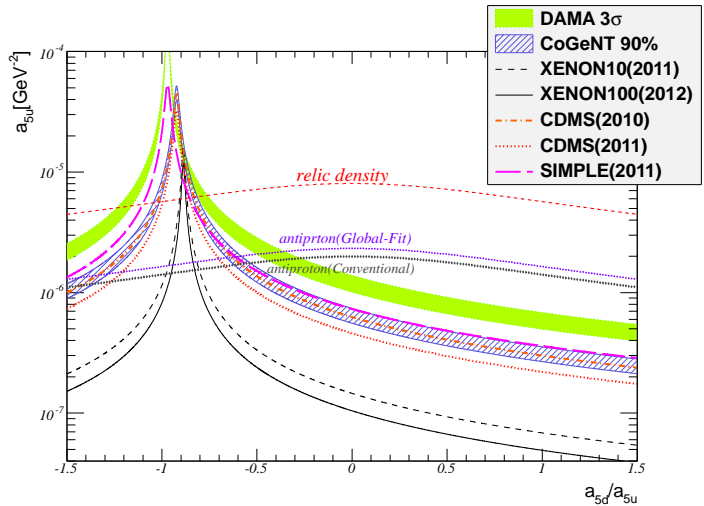}
\end{center}
\caption{
  Upper bounds on the coefficient $a_{5u}$ as a function of $a_{5d}/a_{5u}$ at $95\%$ CL
  from cosmic antiproton flux and DM relic density. The mass of DM particle is fixed at 8 GeV.
  The favored regions and exclusion contours from various experiments such as DAMA \cite{0808.3607},
  GoGeNT \cite{1106.0650}, XENON \cite{1104.3088,1207.5988}, CDMS \cite{1010.4290,1011.2482} and SIMPLE \cite{1106.3014} are also shown.
  }
\label{fig:vector}
\end{figure}

For the operators contribute to $s$-wave annihilation which leads to
velocity-independent annihilation cross sections, the relevant DM couplings to
quarks are found to be tightly constrained by both the cosmic antiproton flux
and the thermal relic density.  In Fig.~\ref{fig:vector} the constraints on
the coefficients $a_{5q}$ for operator
$\mathcal{O}_{5q}=\bar{\chi}\gamma^{\mu}\chi\bar{q}\gamma_{\mu}q$ are shown in
the $(a_{5u}, a_{5d}/a_{5u})$ plane. The mass of the Dirac DM particle is
fixed at $m_{\chi}=8$ GeV.  The results show that the DAMA- and CoGeNT-favored
regions are in tension with both the cosmic antiproton flux and the
thermal relic density.

\begin{figure}[htb]
\begin{center}
 \includegraphics[width=0.48\textwidth]{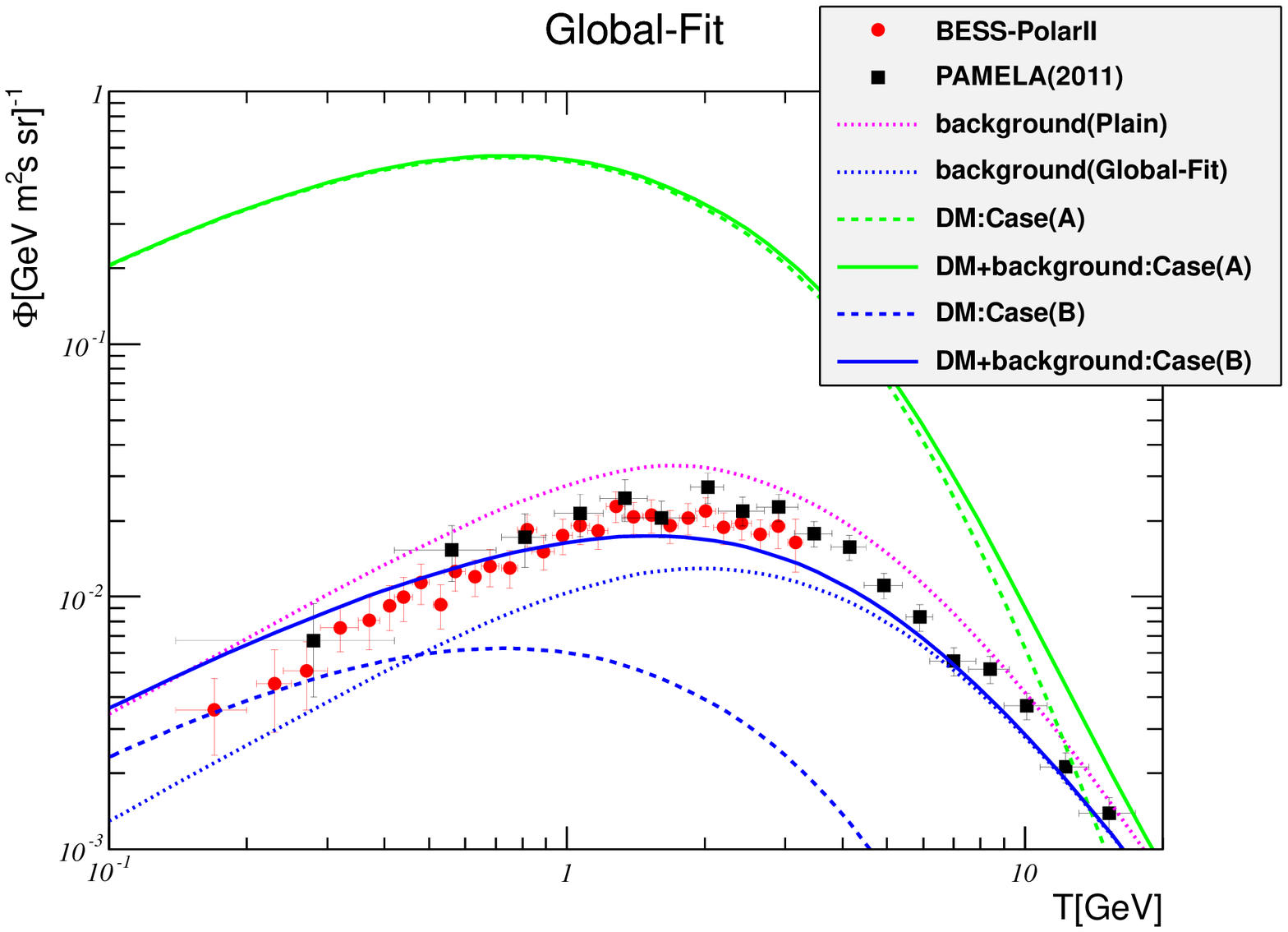}
\includegraphics[width=0.48\textwidth]{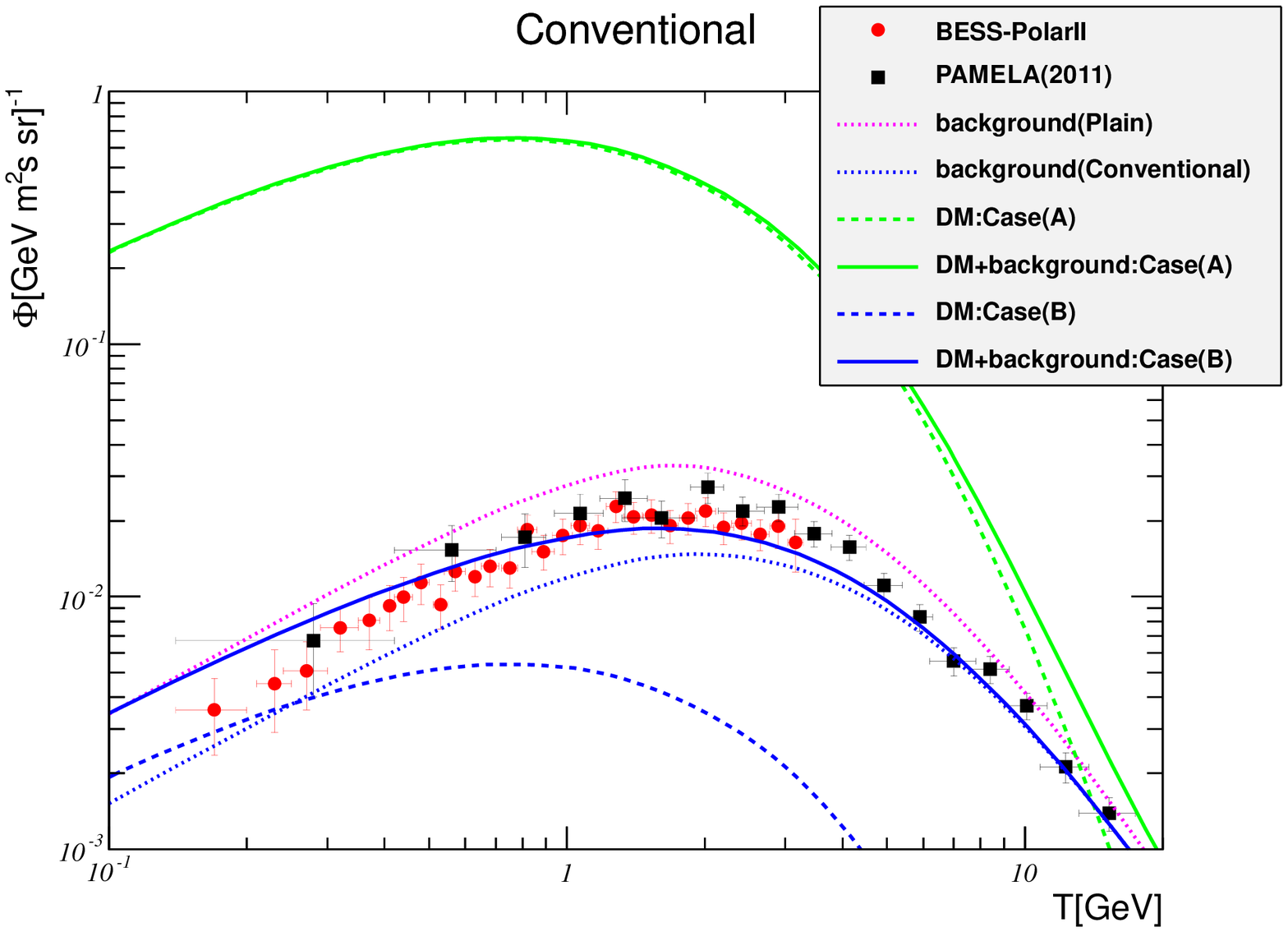}
\end{center}
\caption{ Left) predictions of cosmic antiproton spectra from DM annihilation
  induced by operator $\mathcal{O}_{5q}$ in the ``Global-Fit'' propagation
  model.  Two cases are considered: (A) For $a_{5u}=1.6\times 10^{-5} \text{
    GeV}^{-2}$ which is favored by the DAMA and CoGeNT experiments. (B) For
  $a_{5u}=1.7\times 10^{-6} \text{ GeV}^{-2}$ which is the maximal value allowed by the
  cosmic antiproton data at $95\%$ CL. The ratio $a_{5d}/a_{5u}$ is fixed at
  $-0.89$ corresponding to $f_n/f_p=-0.70$ and the mass of DM particle is fixed at
  8 GeV. The data of BESS-Polar II \cite{1107.6000} and PAMELA \cite{1103.2880} are also shown;
  right )  Same as left), but  for the ``Conventional'' model. }
\label{fig:vector-pbar}
\end{figure}

At $a_{5d}/a_{5u}=-0.89$ which corresponds to $f_n/f_p=-0.70$, the DAMA and
CoGeNT favored value is $a_{5u} \approx 1.6\times 10^{-5} \text{ GeV}^{-2}$
corresponding to an annihilation cross section of $\langle \sigma v_{\text{rel}} \rangle \approx 3.4\times 10^{-25}
\text{ cm}^{2}$.  As it can be seen in Fig. \ref{fig:vector-pbar}, for such a
large cross section, the predicted antiproton flux is much higher than the
current BESS-Polar II and PAMELA data and results in a  huge
$\chi^2/$d.o.f=$1.3\times 10^{6}/35$  in the ``Global-Fit''  model and an even larger one in the ``Conventional'' model.
The upper bound set by the antiproton data at $95\%$ CL is $a_{5u} \leq 1.7\times 10^{-6} \text{ GeV}^{-2}$ in  the
``Global-Fit '' model at $a_{5d}/a_{5u}=-0.89$, which is about an order of magnitude lower and
corresponds to an annihilation cross section $\langle \sigma v_{\text{rel}} \rangle \approx
3.7\times 10^{-27}  \text{cm}^{2}$. The predicted antiproton fluxes are also shown in Fig.
\ref{fig:vector-pbar}.
In the two  propagation models, the upper bound from the ``Conventional'' model is slightly
stronger than that from the ``Global-Fit'' model.  In
Fig. \ref{fig:vector-pbar}, we also plot the antiproton background of
``Plain'' model which is already higher than the data. Thus the upper bounds
from this propagation model are expected to be  much  stronger. Since we are interested in
conservative upper bounds, we do not further investigate  the constraints in this model.
In Fig. \ref{fig:vector-pbar}, the upper bound from the relic density is
$a_{5u} \leq 6.0\times 10^{-6} \text{ GeV}^{-2}$ at $a_{5d}/a_{5u}=-0.89$,
which is weaker than that from the antiproton flux but still in tension with
the DAMA- and CoGeNT-favored value.

\begin{figure}[htb]
\begin{center}
  \includegraphics[width=0.6\textwidth]{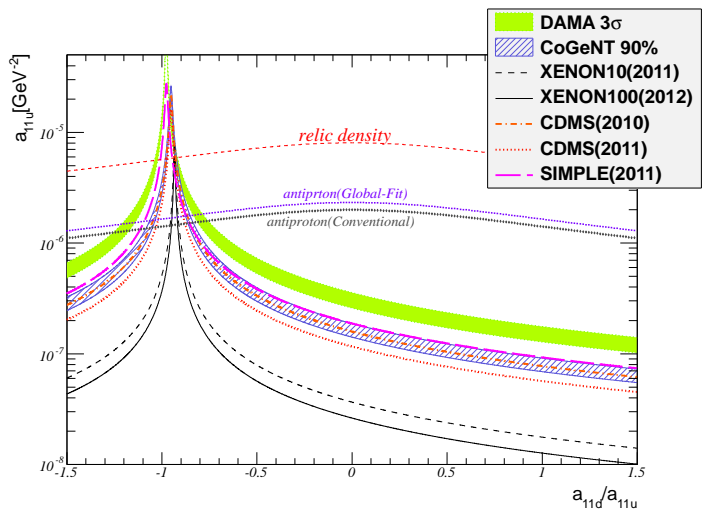}
\end{center}
\caption{Same as Fig. \ref{fig:vector}, but for coefficients $a_{11q}$ of operator $\mathcal{O}_{11}=2m_{\phi}(\phi^{*}\phi) (\bar{q}q)$.
}
\label{fig:complex-scalar}
\end{figure}

\begin{figure}[htb]
\begin{center}
 \includegraphics[width=0.48\textwidth]{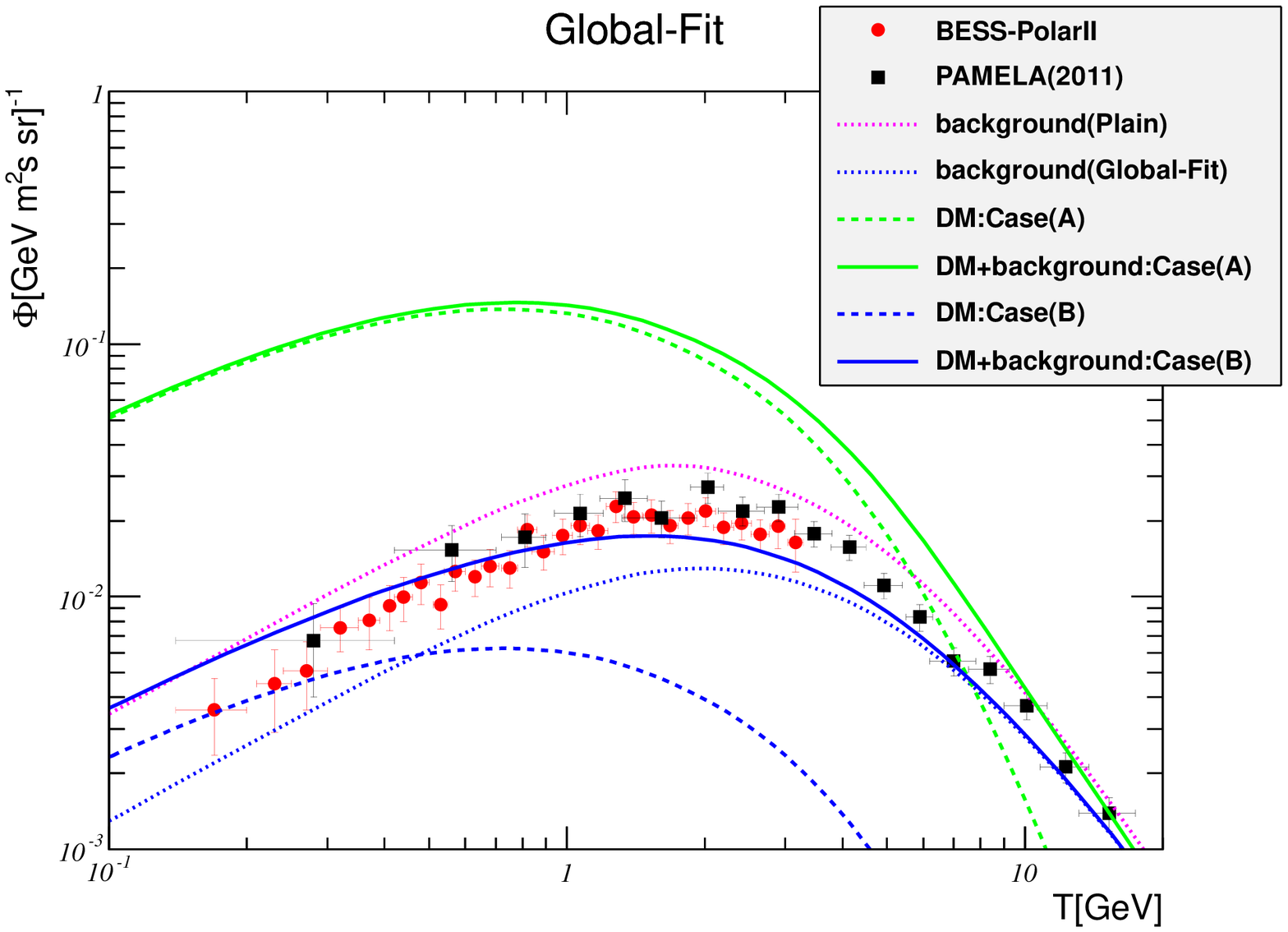}
\includegraphics[width=0.48\textwidth]{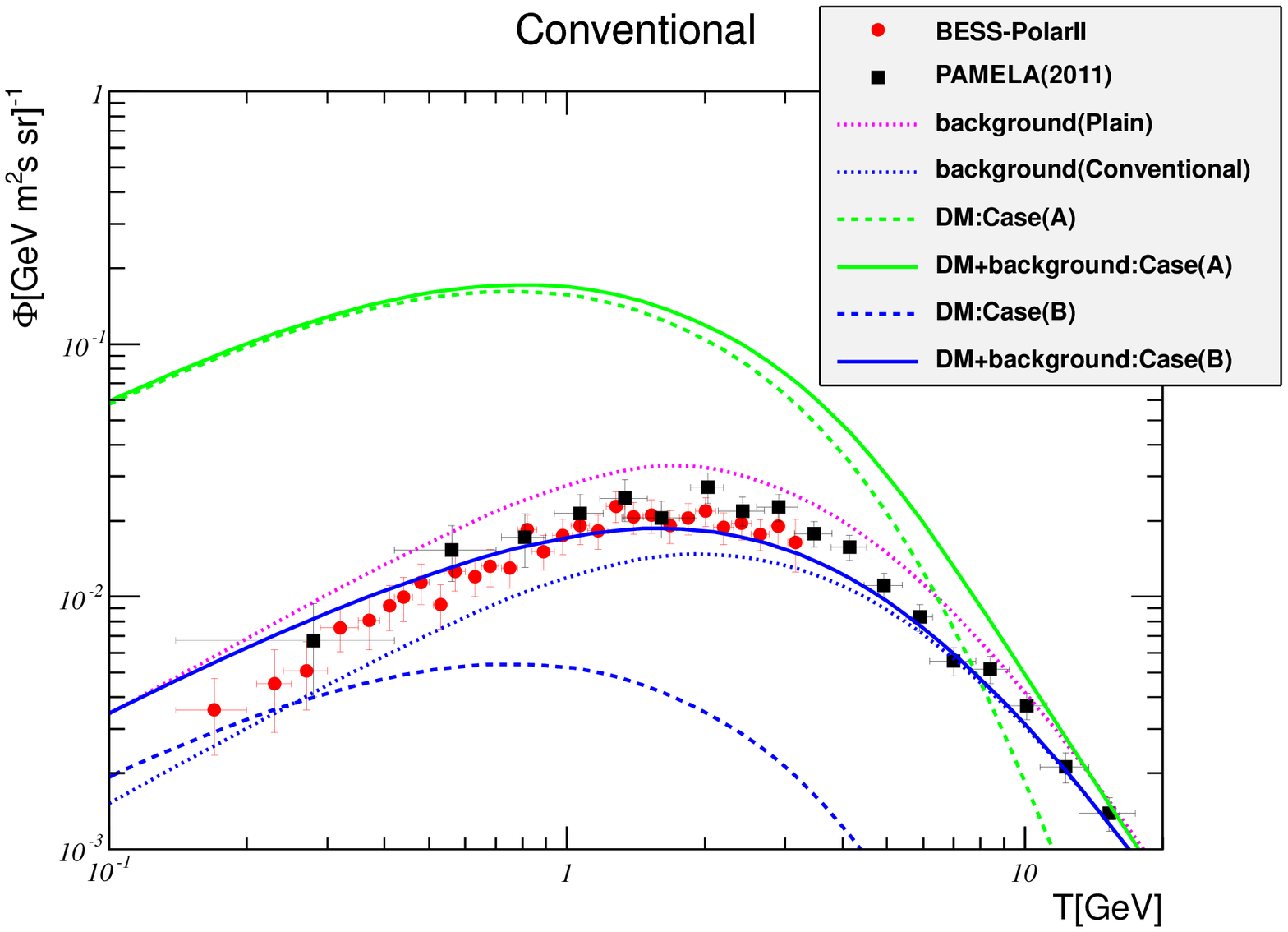}
\end{center}
\caption{  Left) predictions of cosmic antiproton spectra from DM annihilation
  induced by operator $\mathcal{O}_{11q}$ in the ``Global Fit'' propagation
  model.  Two cases are considered: (A) For $a_{11u}=7.9\times 10^{-6} \text{
    GeV}^{-2}$ which is favored by the DAMA and CoGeNT experiments. (B) For
  $a_{11u}=1.7\times 10^{-6} \text{ GeV}^{-2}$ which is the maximal value allowed by the
  cosmic antiproton data at $95\%$ CL. The ratio $a_{11d}/a_{11u}$ is fixed at
  $-0.93$ corresponding to $f_n/f_p=-0.70$ and the mass of DM particle is
  8 GeV. The data of BESS-Polar II \cite{1107.6000} and PAMELA \cite{1103.2880} are also shown;
  right )  same as left), but  for the ``Conventional'' model. }
\label{fig:complex-scalar-pbar}

\end{figure}

In Fig. \ref{fig:complex-scalar} we show the
constraints on the coefficients $a_{11q}$ for operator
$\mathcal{O}_{11}=2m_{\phi}(\phi^{*}\phi) (\bar{q}q)$ in $(a_{11u}, a_{11d}/a_{11u})$
plane for the complex scalar DM mass fixed at $m_\phi=8$ GeV.
For this operator, the DAMA- and CoGeNT-favored regions are still
in tension with the antiproton flux.
At $a_{11d}/a_{11u}=-0.93$ which corresponds to $f_{n}/f_{p}=-0.70$, the DAMA- and
CoGeNT-favored values is $a_{11u} \approx 7.9\times 10^{-6} \text{ GeV}^{2}$, corresponding to an annihilation
cross section of $\langle \sigma v_{\text{rel}} \rangle \approx 8.5\times 10^{-26} \text{ cm}^{2}$. The predicted antiproton fluxes
are shown in Fig. \ref{fig:complex-scalar-pbar} which is again much higher than the current data. The
upper bound set by the antiproton data is $a_{11u} \leq 1.7\times 10^{-6} \text{ GeV}^{2}$ at $a_{11d}/a_{11u}=-0.93$ .
Compared with
the case of $\mathcal{O}_{5q}$, the constraints on  the coefficients of  $\mathcal{O}_{11q}$ are weaker, which  is
due to the fact that for the hadronic matrix element of scalar operator $\bar{q}q$ the $B_{iq}$ factors are larger
than that for vector operator $\bar{q}\gamma^{\mu}q$, which  allows smaller
$a_{iq}$ for the same value  of $f_{p,n}$ and results in  smaller annihilation cross sections. The
constraint  from the thermal relic density is in tension with the DAMA- and
CoGeNT-favored values.

\begin{figure}[htb]
\begin{center}
  \includegraphics[width=0.60\textwidth]{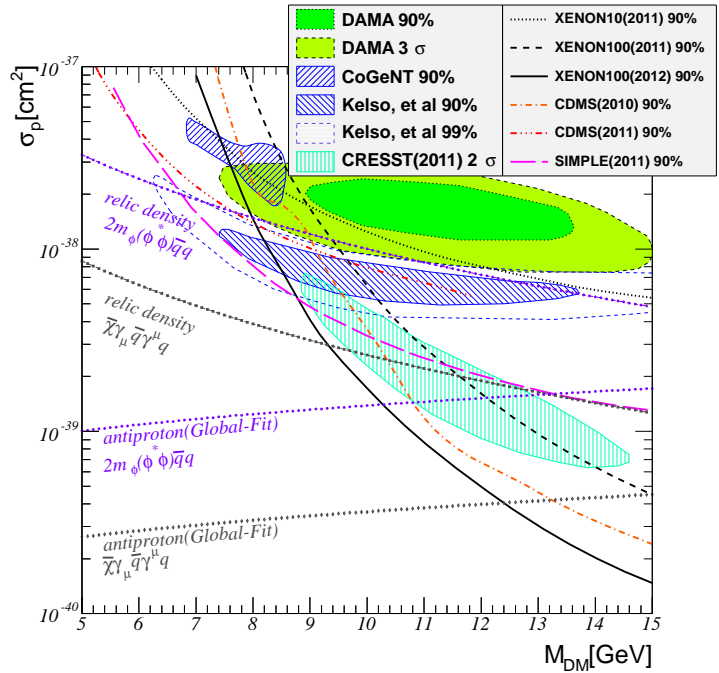}
\end{center}
\caption{The favored regions and constraints in the $(\sigma_p,
  m_{DM})$ plane for operators  $\mathcal{O}_{5q}$ and $\mathcal{O}_{11q}$ with
  the coefficients corresponding to $f_{n}/f_{p}=-0.7$.}
\label{fig:cross-section}
\end{figure}

In Fig. \ref{fig:cross-section}, we show the favored regions and exclusion
contours for operator $\mathcal{O}_{5q}$ and $\mathcal{O}_{11q}$ for the
coefficients corresponding to $f_n/f_p=-0.70$ in $(\sigma_p, m_{DM})$ plane
where $m_{DM}$ is the mass of the  Dirac or complex scalar DM particle in the range
from 5 GeV to 15 GeV. For both operators, the DAMA- and CoGeNT-favored
regions are far above the $95\%$ CL bounds from the antiproton flux for all the values of DM
mass in this range. The favored regions are also in tension with the
thermal relic density, especially for $\mathcal{O}_{5q}$.

\begin{figure}[htb]
\begin{center}
  \includegraphics[width=0.45\textwidth]{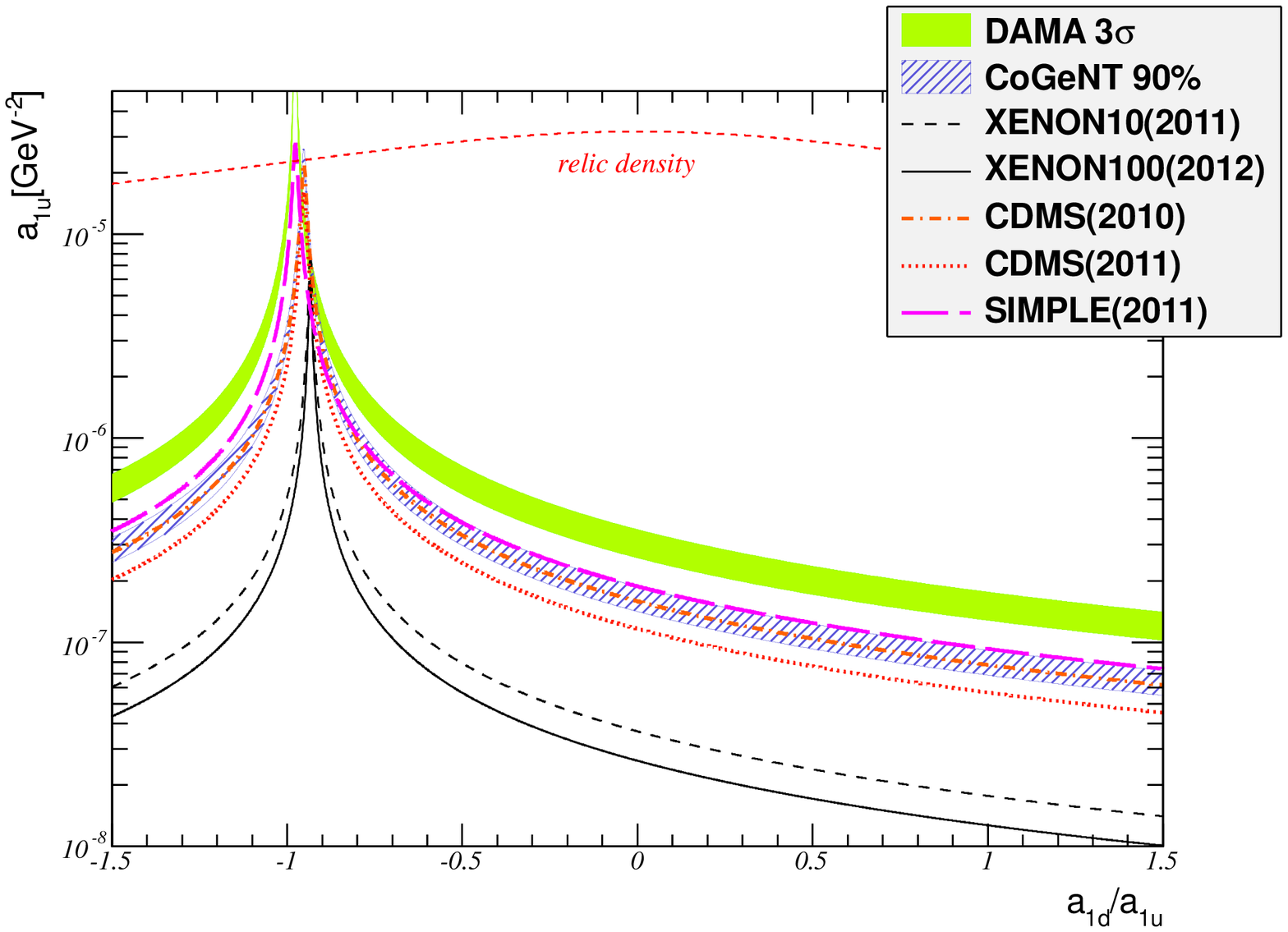}
   \includegraphics[width=0.45\textwidth]{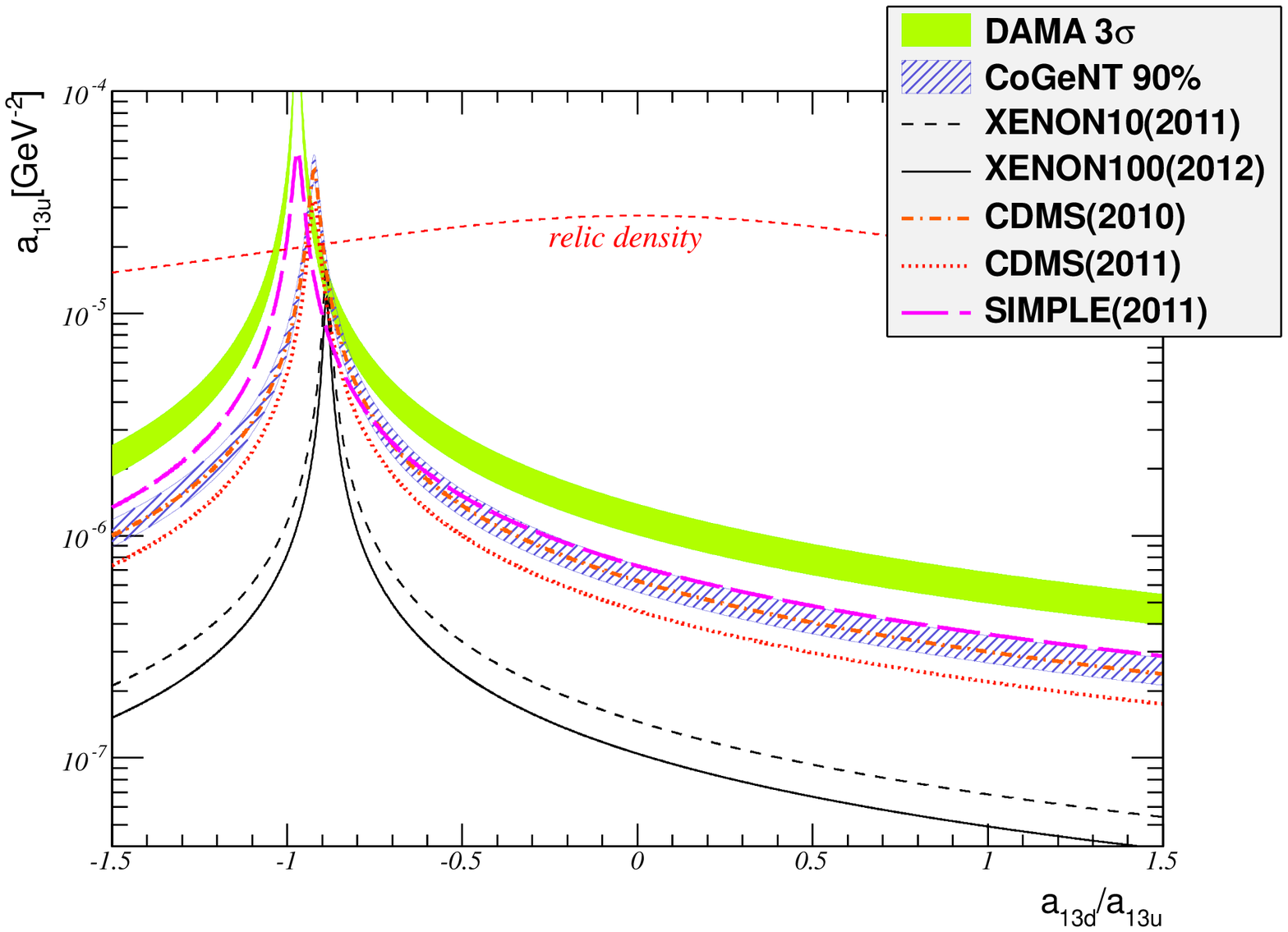}
\end{center}
\caption{Left) constraints from thermal relic
  density on the coefficients $a_{1q}$ of operator $\mathcal{O}_{1q}=\bar{\chi}\chi\bar{q}q$
  in the $(a_{1u}, a_{1d}/a_{1u})$ plane.  The mass of  Dirac DM particle is fixed at $m_{\chi}=8$ GeV.  The
  favored regions and exclusion contours from various experiments are also shown;
  right) same as left), but for coefficients $a_{13q}$.}
\label{fig:p-wave}
\end{figure}

The operators $\mathcal{O}_{1q}= \bar{\chi}\chi\bar{q}q$ and
$\mathcal{O}_{13q}=(\phi^{*}\overleftrightarrow{\partial_{\mu}}\phi )\bar{q}
\gamma^{\mu} q $ contribute to $p$-wave annihilation with cross section
proportional to $v_{\text{rel}}^{2}$. Thus they do contribute very little to
the cosmic antiproton flux. However, their contributions to the thermal relic
density cannot be neglected, as at freeze out the relative velocity is finite.
In Fig. \ref{fig:p-wave}, we show the constraints from relic density on the
coefficients $a_{1q}$ and $a_{13q}$. For the operator $\mathcal{O}_{13q}$ one
can see some tension between bounds set by the relic density and regions
favored by DAMA and CoGeNT data. The constraint is not as stringent as that from the latest
XENON100 data.  For the operator $\mathcal{O}_{1q}$, the
constraints from relic density is rather weak. The difference is again due to
the different $B_{iq}$ factors for these two type of interactions.

It is straight forward to extend the discussions to Majorana fermions and real
scalars. For the particles being its own antiparticles the primary source of
the anitproton will be enhanced by a factor of 2 in Eq. (\ref{eq:ann-source}),
which may lead to more stringent constraints.

\section{Conclusions}\label{sec:conclusions}
In summary, we have investigated the allowed values of DM-nucleon couplings in
the scenario of IVDM for various target nuclei used
in DM  direct detections. We find that the recently updated XENON100
  result  excludes the main part of the overlapping  signal region between
  DAMA and CoGeNT.     We have shown that whereas the effect of
isospin violating scattering can relax the tensions between the data of DAMA,
CoGeNT and XENON, the possible disagreement between some group of experiments
such as that between DAMA  and SIMPLE are not likely to be affected for any value of $f_{n}/f_{p}$.
 In an effective operator approach, we have investigated the
conservative constraints on the couplings between the IVDM and the SM light quarks from the
recent cosmic ray antiproton  data and that
from the thermal relic density. Among  the four operators relevant to IVDM
$\mathcal{O}_{1q}$, $\mathcal{O}_{5q}$, $\mathcal{O}_{11q}$,
$\mathcal{O}_{13q}$,   the operators $\mathcal{O}_{5q}$ and $\mathcal{O}_{11q}$
are found to be tightly constrained by the antiproton data and $\mathcal{O}_{13q}$ is
constrained by the relic density. Only the operator $\mathcal{O}_{1q}$ can survive
both the  constraints while contribute to large enough isospin violating interaction
required by the current data of DAMA, CoGeNT and XENON.

The scenario of IVDM may be less constrained by the cosmic ray observations if
the DM particles interact with SM particles only through some mediator
particle $\phi$ such that the DM annihilation cannot be described by effective
operators. For instance, if the mass of the light mediator $m_{\phi}$ is much
smaller than that of the dark matter particle $m_{\chi}$ while still
significantly larger than the typical recoil energy $\sim \mbox{keV}$, the
effective operator approach is valid only for the elastic scattering but not
for the annihilation. The cross section for DM annihilation will be suppressed
by a factor of $m_{\phi}^{4}/m_{\chi}^{4}$ compared with the ordinary
effective operator approach.  The IVDM is also less constrained by the cosmic
ray data if it is asymmetric.

\subsection*{Acknowledgments}
We are grateful to Yue-Liang Wu for encouragement and helpful discussions.
This work is supported in part by
the National Basic Research
Program of China (973 Program) under Grants No. 2010CB833000;
the National Nature Science Foundation of China (NSFC) under Grants
No. 10975170,
No. 10821504 and No. 10905084;
and the Project of Knowledge Innovation Program
(PKIP) of the Chinese Academy of Science.

% \newpage
\bibliographystyle{JHEP}

\providecommand{\href}[2]{#2}\begingroup\raggedright\endgroup

\end{document}